\documentclass[%
 reprint,
superscriptaddress,
 amsmath,amssymb,
 aps,
]{revtex4-2}

\usepackage{graphicx}% Include figure files
\usepackage{dcolumn}% Align table columns on decimal point
\usepackage{bm}% bold math
\newcommand{\schro}{Schr\"{o}dinger }

\begin{document}

\preprint{APS/123-QED}

\title{Using the Ehrenfest theorem for determining the self-focusing and self-trapping of nonlinear beams}% Force line breaks with \\
%thanks{A footnote to the article title}%

\author{Chandroth P. Jisha}
 \email{jisha.chandroth.pannian@uni-jena.de}
 \affiliation{%
    Friedrich Schiller University Jena,
    Abbe Center of Photonics,  Institute of Applied Physics, Albert-Einstein-Str. 15, 07745 Jena, Germany
}%

%\affiliation{CNR-ISASI, Institute of Applied Science and Intelligent Systems, Via Campi Flegrei 34, 80078 Pozzuoli (NA), Italy}

\author{Stefan Nolte}%
\affiliation{%
    Friedrich Schiller University Jena,
    Abbe Center of Photonics,  Institute of Applied Physics, Albert-Einstein-Str. 15, 07745 Jena, Germany
}%
\affiliation{Fraunhofer Institute for Applied Optics and Precision Engineering IOF, Albert-Einstein-Str. 7, 07745 Jena, Germany}

\author{Alessandro Alberucci}%
\email{alessandro.alberucci@uni-jena.de}
\affiliation{%
  Friedrich Schiller University Jena,
    Abbe Center of Photonics,  Institute of Applied Physics, Albert-Einstein-Str. 15, 07745 Jena, Germany
}%

\date{\today}% It is always \today, today,

\begin{abstract}
We discuss how to generalize the Ehrenfest theorem for the computation of the width of nonlinear waves obeying the nonlinear \schro equation. To do that, we model the nonlinear potential as a quantum harmonic oscillator (QHO) whose strength depends on the power and on the wavefunction width. We apply the model to different types of nonlinear responses, eventually comparing the results with numerical simulations. Our model has the advantage of explaining the main properties of nonlinear confined waves, such as stability and breathing, in  a relatively simple and intuitive manner.   
\end{abstract}

\maketitle

%\tableofcontents

\section{Introductory notes}

Nonlinear effects are common to every field of physics. A nonlinear response is intrinsic to the dynamics of viscous fluids as modeled by the Navier-Stokes equation \cite{Landau:1987},  or to the mechanical deformation of a solid \cite{Landau:2012}. In electromagnetism, the Maxwell's equations become nonlinear when the electromagnetic field is strong enough to induce a nonlinear relation between the applied field and the induced dipoles in the material \cite{Boyd:1992}. In optics, the functioning of basic devices, such as mode-locked lasers, takes place in the nonlinear regime. In BEC, the condensates see a nonlinear response once the interaction between particles is accounted for \cite{Leggett:2001}. 

Perhaps the easier and one of the most common manifestation of nonlinear physics in wave propagation is the Kerr effect, that is, when the local phase of the wave depends on the intensity of the wave in the same point \cite{Kelly:1965}. When the wave in the linear regime is dispersive, the Kerr effect can induce a nonlinearly-induced self-focusing. Once a balance between dispersion and self-focusing is achieved, waves do not modify their shapes in propagation \cite{Chiao:1964}: these are called solitons in the integrable case, solitary waves in the non-integrable case \cite{Dauxois:2006}.

A common and widespread model for the propagation of dispersive waves in the presence of a self-phase modulation is the generalized nonlinear \schro equation (GNLSE) \cite{Malkin:1993,Kartashov:2019}
\begin{equation}
    i\hbar\frac{\partial \psi}{\partial t}=-\frac{\hbar^2}{2m}\nabla^2 \psi + V(\psi) \psi.
\end{equation}
Nonlinearity is then accounted for by a potential $V$ dependent on the field $\psi$ itself.  When the potential is proportional to the field intensity, i.e., $V\propto |\psi|^2$, we obtain the standard NLSE, successfully employed in several fields such as optics, BEC, plasma physics, water waves, molecular excitations \cite{Davydov:1982,Dauxois:2006,Chabchoub:2019}.

To fix the ideas, hereafter we will focus on the optical case considering a paraxial scalar monochromatic wave propagating in space. Noteworthy, the NLSE in optics models also other systems, such as the nonlinear propagation in a fiber or the spatio-temporal propagation of a pulse in the space-time \cite{Kivshar:2003}. When the NLSE is applied to the nonlinear evolution of optical CW (continuous wave) fields, the evolution $t$ becomes the propagation coordinate $z$, whereas the Laplacian is now limited to two transverse coordinates, $\nabla^2_T=\partial_x^2 + \partial_y^2$ \cite{Longhi:2009} and $\bm r_T = x\hat{x} + y\hat{y}$. Calling $n$ the refractive index and $k_0$ the vacuum wave number, the evolution of the optical field is governed by 
\begin{equation} \label{eq:NLSE}
    i\frac{\partial \psi}{\partial z}= -\frac{1}{2k_0 n_0} \nabla_T^2 \psi - \frac{k_0 \Delta n^2(\psi)}{2n_0} \psi,
\end{equation}
where $n_0$ is the unperturbed (i.e., in the linear regime) refractive index, in our case assumed to be uniform across the space. The equivalent mass is then given by $k_0n _0$, whereas $\hbar$ is fixed equal to unity \cite{Cook:1975}.
Given the perturbative character of most of the nonlinear effects, the nonlinear potential is related to the nonlinear change in the refractive index $\Delta n_\mathrm{NL}$ via
\begin{equation} \label{eq:VNL}
    V_\mathrm{NL}(\psi) \approx - k_0 \Delta n_\mathrm{NL}(\psi).
\end{equation}

The propagation of light in a Kerr material has been of uttermost importance since the inception of laser \cite{Boyd:2009}: permanent damages observed in dielectric media can indeed be described as a consequence of catastrophic collapse \cite{Kelly:1965,Akhmanov:1968,Shen:1975}. A vast amount of theoretical work, including numerical simulations and theoretical simplified models, has been then dedicated to explain the experiments \cite{Marburger:1975,Kivshar:2000,Couairon:2007}. In the monochromatic regime, common theoretical approaches include ray-optics models \cite{Lim:2014}, self-similar solutions recalling Gaussian beams in the linear regime \cite{Akhmanov:1968}, the method of moments \cite{Lam:1975}, and the variational approach \cite{Anderson:1979,Malomed:2002,Jisha:2005}. Interestingly, the application of the second-order moment to the conservation laws for an electromagnetic field was already employed by Vlasov in 1971 \cite{Vlasov:1971} to describe the collapse of an optical beam when the power overcomes the critical power $P_\mathrm{cr}$. On the side of nonlocal nonlinear media \cite{Bang:2002}, Snyder and Mitchell introduced in 1997 \cite{Snyder:1997} the concept of accessible solitons as self-confined waves in a power-dependent parabolic potential, later demonstrated experimentally in nematic liquid crystals \cite{Conti:2004}. 

Here, we develop an alternative approach to the moments method using the Ehrenfest theorem in its generalized form, the latter providing the expectation values of any operator in quantum mechanics \cite{Sakurai:1994}. Similarly to the aberration-free approach \cite{Akhmanov:1968,Magni:1993} and the accessible soliton model \cite{Snyder:1997,Conti:2004}, we assume the nonlinear potential to be a quantum harmonic oscillator encompassing an intensity-dependent strength. A nonlinearity-independent criterion for fixing the nonlinear parabolic potential is elaborated. In this limit, we show how the Ehrenfest theorem provides a fourth-order ordinary differential equation (ODE) in the beam width, generalizing a result for the highly nonlocal case demonstrated in Ref.~\cite{Alberucci:2016_1}. We then show how this equation can be transformed into a second-order ODE with appropriate boundary conditions, the latter dependent on the launch conditions. We prove the versatility and simplicity of our approach by modeling different nonlinear responses, including local and nonlocal materials, higher-order Kerr effects, and cubic-quintic media. Comparison with other theoretical methods, such as the variational approach, and numerical simulations based upon a beam propagation method (BPM) is provided as well. 

The Article is structured as follows. Section~\ref{sec:Ehrenfest} contains the core theoretical results of the paper, culminating in Eq.~\eqref{eq:master_equation_with_boundary}: we show how, using the Ehrenfest's theorem and assuming a parabolic nonlinear potential, a single ODE equation for modelling the propagation of waves subject to self-focusing can be derived. In Section~\ref{sec:QHO} we derive the equivalent QHO for a set of well-known nonlinear materials in optics (Kerr, cubic-quintic, nonlocal). In Section~\ref{sec:applications} we present the applications of our theoretical findings to real cases, eventually comparing our results with numerical simulations. In Section~\ref{sec:conclusions} we summarize our results, discuss their relevance and providing a brief perspective on future generalizations.

\section{Application of the Ehrenfest theorem to a nonlinear quantum harmonic oscillator}
\label{sec:Ehrenfest}
In quantum mechanics, an evolution equation for any operator $\hat{A}$ can be found using the commutator with the Hamiltonian $\hat{H}$ of the system \cite{Sakurai:1994,Styer:1990}, in our optical framework reading $d\hat{A}/dz = i[\hat{A},\hat{H}]+\partial \hat{A}/\partial z$. For $\hat{A}=\hat{x}$ and $\hat{A}=\hat{p}$, the standard Ehrenfest theorem can be found, that is, the centroid or first moment $\langle \bm r_T \rangle=  \int{\bm r_T |\psi|^2dxdy}\left/{\int{|\psi}|^2dxdy}\right.$ of the wave packet moves in a potential according to the classical Newton law. The Ehrenfest theorem for the beam trajectory has been already extensively used in the case of spatial optical solitons \cite{Alfassi:2006,Alberucci:2007_2,Jisha:2011,Garza:2019} or for beams propagating in a random material \cite{Cook:1975}. 

For our purposes, we introduce the beam width $w$ as proportional to the second moment of the position $w_x^2 = 4\langle x^2 \rangle$ and $w_y^2 = 4\langle y^2 \rangle$. With this choice, $w_x$ and $w_y$ correspond to the width $w$ of a Gaussian beam as usually defined in optics through the position $I\propto \exp{(-2x^2/w^2)}$. To avoid the usage of centered moments and simplify the notation as much as possible, hereafter we assume that the beam is always placed at the origin $(x=0,y=0)$. The application of the Ehrenfest theorem to $\hat{\bm r}_T^2 $ and $\hat{\bm p}^2$ provides a system of two second-order equations for the second-order moments of the position and of the momentum, namely $\langle \bm r_T^2 \rangle$ and $\langle \bm p^2 \rangle$ \cite{Alberucci:2016_1,Hansson:2020}. When the moment equations are found by means of a variational approach \cite{Caglioti:1990}, a system of two coupled ODE -one for the beam size and one for the phase- is found, thus confirming the physical equivalence of the two approaches.
%The equation governing the beam width evolution can be derived using the analogy between paraxial beam propagation and quantum mechanics.

Hereafter, we will limit our discussion to the radially-symmetric case. After defining the radial distance $r_T=\sqrt{x^2+y^2}$, we suppose the presence of a parabolic potential in the form \cite{Belanger:1983}
\begin{equation} \label{eq:parabolic_approximation}
    V_\mathrm{NL}(r_T,z) = k_0 \frac{a(z)}{2} r_T^2, 
\end{equation}
where in our case the strength of the QHO depends on the propagation distance $z$ due to the variations in propagation of the wave $\psi$. In quantum mechanics the parabolic position ascertained by Eq.~\eqref{eq:parabolic_approximation} corresponds to the so-called local harmonic oscillator \cite{Heller:1975} which provides a set of semi-classical equations of motion for the wave packet.  A self-focusing nonlinearity requires $a>0$. 
  
In the case of a parabolic potential, a single fourth-order equation can be written down: the square of the beam width $w$ indeed evolves along the propagation coordinate $z$ according to a fourth-order ODE \cite{Alberucci:2016_1}
\begin{equation} \label{eq:ehrenfest_harmonic}
    \frac{n_0}{2} \frac{d^4 w^2}{dz^4} + 2a \frac{d^2 w^2}{dz^2} + 3 \frac{da}{dz} \frac{dw^2}{dz} + \frac{d^2 a}{dz^2} w^2 =0.
\end{equation}
From Eq.~\eqref{eq:ehrenfest_harmonic} it is evident that a self-consistent model can be found by assuming that the QHO strength $a$, beyond the power $P$, depends on the beam width $w$. On the other side, Kerr-like nonlinear effects depend on the local intensity of the wave $I$. With respect to $P$ and $w$, we can set $I\propto P/w^2$. These considerations together support the generalized ansatz
\begin{equation} \label{eq:ansatz_a}
    %a(z,P) =  \gamma(w) \left[\frac{ P(z)}{w^2(z)}\right]^N.
     a(z,P) =  \gamma \frac{ P^N(z)}{w^{2M}(z)},
\end{equation}
where in general $N\neq M$. In the case of a nonlinearity of $L$-th order, from $V_\mathrm{NL}\propto I^{(L-1)/2}$ we get $2N+1=L$. We will show later that the counter-intuitive condition $N\neq M$ for some nonlinear responses is owed to a proper averaging of the nonlinear transverse gradient across the wave cross-section. \\
The quantity $\gamma$ depends on the magnitude and type of nonlinearity considered, whereas $N$ and $M$ depend on the type of nonlinearity considered.   In the simultaneous presence of multiple nonlinearities (for example, $V_\mathrm{NL}(I)=-k_0\sum_m n_{2m} I^m$ in the presence of higher-order Kerr effect \cite{Bejot:2011}), Eq.~\eqref{eq:ehrenfest_harmonic} predicts that the strength of the harmonic oscillator $a$ is given by the sum of the contribution from each component.
For compactness, we introduce the auxiliary quantity $\kappa(z) = \gamma P^N$. Power $P$ can change in propagation in the case, for example, of dissipative systems: well-known examples are ultrashort pulses in the presence of multi-photon ionization (MPI), thermo-optical materials \cite{Dabby:1968,Alfassi:2006}, or nematic liquid crystals (NLCs) subject to strong scattering losses \cite{Alberucci:2018}.

Substituting Eq.~\eqref{eq:ansatz_a} into Eq.~\eqref{eq:ehrenfest_harmonic} yields
%\begin{equation} \label{eq:equation_fourth_order_intermediate}
%  \frac{d^4 w^2}{dz^4} + \frac{2(2-N)\kappa}{n_0}\left[\frac{1}{w^{2N}  } \frac{d^2 w^2}{dz^2} - \frac{N}{w^{2(N+1)}} \left(\frac{dw^2}{dz} \right)^2 \right]=0,
%\end{equation}
\begin{equation} \label{eq:equation_fourth_order_intermediate}
  \frac{d^4 w^2}{dz^4} + \frac{2(2-M)\kappa}{n_0}\left[\frac{1}{w^{2M}  } \frac{d^2 w^2}{dz^2} - \frac{M}{w^{2(M+1)}} \left(\frac{dw^2}{dz} \right)^2 \right]=0,
\end{equation}
where we neglected the derivatives of the power with respect to $z$, that is, we assumed an adiabatic drop in the optical power (see Appendix~\ref{sec:der_power} for the general case). 
For $M=2$, it is $d^4 w^2/dz^4=0$. For $M=1$ and noticing that $\partial_z^2 \log w^2=w^{-2}\partial_z^2 w^2 - w^{-4} \left(\partial_z w^2 \right)^2$, we retrieve the highly nonlocal case we previously treated in Ref.~\cite{Alberucci:2016_1}. 

Given that for a generic positive integer $Q\neq 1$
\begin{equation}
    \frac{d^2}{dz^2}\left(\frac{1}{w^{2Q}} \right) = Q \left[ \frac{Q+1}{w^{2(Q+2)}} \left(\frac{dw^2}{dz} \right)^2 - \frac{1}{w^{2(Q+1)}}\frac{d^2 w^2}{dz^2} \right],
\end{equation}
Eq.~\eqref{eq:equation_fourth_order_intermediate} can be recast as
\begin{equation} \label{eq:equation_fourth_order}
    \frac{d^2}{dz^2} \left[ \frac{d^2 w^2}{dz^2} + \frac{2\kappa (M-2)}{n_0 (M-1)} \frac{1}{w^{2(M-1)}} \right] =0.
\end{equation}
We consider a nonlinear material starting from $z=0$, the left side being filled with air.
In agreement with the most common experimental setups, we take a Gaussian beam of waist $w_0$ and focal position $z_0$, thus featuring a Rayleigh length $L_R=k_0 n_0 w_0^2/2$. Then, at the entrance interface ($z\rightarrow 0^-$) we obtain a beam of width $w^2_\mathrm{in}=w_0^2\left[ 1+ (z_0/L)^2 \right]$  and curvature radius $R_0= z_0 + L_R^2/z_0$.

Equation~\eqref{eq:equation_fourth_order} is an initial value problem: it must be solved jointly with the boundary conditions for $w^2$ and its derivatives versus $z$ up to the third order. 
Given our launch conditions, it is straightforward to get $ \left.{w}^2\right|_{z=0} = w^2_\mathrm{in}$.
For the first derivative we need to consider the general definition of curvature radius $R^{-1}\equiv [1/(2w^2)dw^2/dz]$ \cite{Siegman:1991,Belanger:1983}, showing how $R$ determines the variations along $z$ of the beam width. 
To apply such a definition, we consider at the interface the overall curvature radius will be dictated by the sum of the impinging phase with the nonlinearly-induced phase profile stemming from $V_\mathrm{NL}$ \cite{Belanger:1983}
\begin{equation} \label{eq:boundary_R}
    \left.\frac{1}{R}\right|_{z=0^+} =\frac{1}{R_\mathrm{in}}= \frac{1}{R_\mathrm{0}} + \frac{1}{R_\mathrm{NL}} ,
\end{equation}
in turn providing the following condition for the first derivative
\begin{equation}
    \left.\frac{dw^2}{dz}\right|_{z=0^+}  = \left.\frac{2w^2}{R} \right|_{z=0^+} . \label{eq:initial_curvature_radius}
\end{equation}
Under the parabolic approximation for the nonlinear potential (see Eq.~\eqref{eq:parabolic_approximation}) immediately follows $R_\mathrm{NL}=k_0 n_0/a$.
In Eq.~\eqref{eq:boundary_R} the curvature radius $R_\mathrm{0}$ is computed on the right side of the interface in the linear regime, that is, at low input powers. \\
Ref.~\cite{Alberucci:2016_1} finally provides the other two remaining conditions for the second and third derivative 
\begin{align}
    %\left.{w}^2\right|_{z=0} &= w^2_\mathrm{in}, \\
    %  \left.\frac{dw^2}{dz}\right|_{z=0}  &= \left.\frac{dw^2}{dz} \right|_{z=0} , \\
    \left.\frac{d^2w^2}{dz^2}\right|_{z=0^+} &= \frac{8\langle p^2 \rangle_\mathrm{in}}{k_0^2 n_0^2}  - \frac{2a(0)w^2_\mathrm{in}}{n_0} , \label{eq:gen_BC_der2}\\
     \left.\frac{d^3w^2}{dz^3}\right|_{z=0^+} & = -\frac{4a(0)}{n_0} \left.\frac{dw^2}{dz}\right|_{z=0^+} - \frac{2\dot{a}(0) w_\mathrm{in}^2}{n_0}. \label{eq:gen_BC_der3}
\end{align}
%Equation~\eqref{eq:gen_BC_der2} hides a subtlety: the second moment of the momentum $\langle p^2 \rangle$ is not computed in the linear regime (i.e., $P\rightarrow 0$), but needs to account for the discontinuity in the curvature radius ascribable to the sudden jump in the nonlinear coefficients (that is, $n_{2m}$ for a generalized local Kerr medium). To account for that, we introduce the beam waist $w_{0P}=w_0(P)$, physically corresponding to the beam waist when only a infinitesimally-thin sheet of nonlinear material is present in $z=0$. Details are provided in Appendix~\ref{sec:NL_boundaries}, where it is shown that $w_\mathrm{0P}\approx w_0$ for the range of powers where our theoretical model works. \\
%Remembering that $\langle p^2 \rangle=1/w_0^2$ independently from $z$ for a Gaussian beam and making use of Eq.~\eqref{eq:ansatz_a},
Using Eq.~\eqref{eq:ansatz_a} and the definition of $\kappa$, the boundary conditions can be recast as follows
\begin{align}
    %\left.{w}^2\right|_{z=0} &= w^2_\mathrm{in}, \label{eq:initial_width} \\
     % \left.\frac{dw^2}{dz}\right|_{z=0}  &= \left.\frac{2w^2}{R} \right|_{z=0} , \label{eq:initial_curvature_radius} \\
    \left.\frac{d^2w^2}{dz^2}\right|_{z=0^+} &= \frac{8\langle p^2 \rangle_\mathrm{in}}{k_0^2 n_0^2}  - \frac{2\kappa}{n_0 w_\mathrm{in}^{2(M-1)}} , \label{eq:bc_der2} \\
     \left.\frac{d^3w^2}{dz^3}\right|_{z=0^+} & =  \frac{2\kappa (M-2)}{n_0 w_\mathrm{in}^{2(M-2)}} \left.\frac{dw^2}{dz}\right|_{z=0}. \label{eq:bc_der3}
\end{align}

On the other side, integrating  Eq.~\eqref{eq:equation_fourth_order} twice we find
\begin{equation} \label{eq:master_equation}
     \frac{d^2 w^2}{dz^2} + \frac{2\kappa (M-2)}{n_0 (M-1)} \frac{1}{w^{2(M-1)}}  = Az + B.
\end{equation}
By direct comparison of Eq.~\eqref{eq:master_equation} with Eqs.~\eqref{eq:bc_der2} and \eqref{eq:bc_der3}, we can now determine the integration constants $A$ and $B$, both of them dependent on the beam power, in the general case. Derivation of Eq.~\eqref{eq:master_equation} yields $A=0$ once  Eq.~\eqref{eq:bc_der3} is accounted for. Physically speaking, if $A$ does not vanish, for long enough $z$ the curve of the beam width versus $z$ will always become concave, thus making self-localization impossible for any type and magnitude of nonlinearity. 
To determine $B$, we need the initial condition for the second derivative of $w^2$. Substituting Eq.~\eqref{eq:bc_der2} into Eq.~\eqref{eq:master_equation} sampled in $z=0$ we find %$B=8/(k_0^2n_0^2 w_{0P}^2) - 2\kappa /(n_0 w_\mathrm{in}^{2(M-1)})$.
$B={8\langle p^2 \rangle_\mathrm{in}}/\left({k_0^2 n_0^2}\right)- 2\kappa \left/\left[n_0(M-1) w_\mathrm{in}^{2(M-1)}\right]\right.$.
By plugging back $A$ and $B$ in the master equation \eqref{eq:master_equation}, we find the final form of the equation that rules the evolution of the beam width
\begin{equation} \label{eq:master_equation_with_boundary}
     \frac{d^2 w^2}{dz^2} + \frac{2\kappa (M-2)}{n_0 (M-1)} \frac{1}{w^{2(M-1)}}  =  \frac{8\langle p^2 \rangle_\mathrm{in}}{k_0^2 n_0^2} %\frac{8}{k_0^2 n_0^2 w_{0P}^2} 
     - \frac{2\kappa}{n_0 (M-1)w_\mathrm{in}^{2(M-1)}}.
\end{equation}
In the limit of low powers, that is, $\kappa\rightarrow 0$, we retrieve the linear propagation of a fundamental Gaussian beam equation encompassing $d^2w^2/dz^2=8/(k_0^2 n_0^2 w_0^2)$, independently of the position of the focus. In the generic nonlinear case, the initial convexity of the beam width is determined by the interplay between the nonlinear lens and the focal position $z_0$ in the linear regime, the latter intervening in the interplay through the beam width at the input interface $w_\mathrm{in}$. %, the focal position in the linear regime contributes to determine via the nonlinear lens.
%Equation~\eqref{eq:equation_fourth_order} finally provides

Equation~\eqref{eq:master_equation_with_boundary} -a generalization of our previous result valid for $M=1$ only \cite{Alberucci:2016_1}- is the main result of the paper. The dynamics of a wave propagating in a nonlinear quantum harmonic oscillator can be described as a nonlinear single second-order ordinary differential equation. Physically speaking, the convexity of $w(z)$ is determined by the interplay between diffraction and self-phase modulation.

\subsection{Particle-like model}

Similarly to what done with the variational approach, the evolution of the beam width can be depicted like a mechanical system evolving versus the effective time $z$ with state variables $q=w^2$ and $p= \dot{q}$ corresponding to a generalized coordinate and its associated momentum, respectively \cite{Henz:1996}. For $M \neq 2$ and $M \neq 1$, the effective particle moves under the influence of a nonlinear potential $U$ 
\begin{equation} \label{eq:particle_potential}
   U(w^2) =  \frac{2\gamma P^N }{n_0 (1-M)} \frac{1}{w^{2(M-2)}} - Bw^2.
\end{equation}
For $M=2$, the potential is strictly linear $U=-Bw^2$, with a slope $B$ independent from the beam width $w$ and determined by the input power $P$, initial width $w_\mathrm{in}$ and linear waist $w_0$. Self-trapping is then intrinsically unstable: either diffraction or catastrophic collapse will occur. In the next section we will indeed prove that this corresponds to a Kerr material, its nonlinear dynamics being well known since the early days of nonlinear optics \cite{Kelly:1965,Vlasov:1971}. When the Kerr effect is accompanied by a higher order (i.e., $N>2$) nonlinearity of defocusing character, a relative minimum can then appear in the potential expressed by Eq.~\eqref{eq:particle_potential}, allowing the existence of stable self-trapped waves in the form of shape-preserving spatial solitons and breathing solitons \cite{Henz:1996}. \\ %Finally, Eq.~\eqref{eq:particle_potential} holds valid even when $N=(j-1)/2$, with $j=0,1,2,...$: for even $j$  \\
Eq.~\eqref{eq:particle_potential} does not hold valid for $M=1$, the latter corresponding to the HNL case when $V_\mathrm{NL}$ satisfies a Poisson equation in the form $\nabla_T^2 \Delta n_\mathrm{NL} = -n^\prime_2I$~\cite{Alberucci:2016_1}. The particle potential $U$ reads
 \begin{equation} \label{eq:particle_potential_HNL}
     U(w^2) = \frac{2n^\prime_2 P}{\pi n_0} w^2 \left( \ln\frac{w^2}{w^2_\mathrm{av}} -1 \right),
 \end{equation}
where $w^2_\mathrm{av}=w^2_\mathrm{in}\exp{\left(w^2_\mathrm{sol} /w^2_\mathrm{in} -1 \right)}$ is the average beam width in propagation, $w_\mathrm{sol}=\sqrt{4\pi/(n_0 n^\prime_2k_0^2 P)}$ the soliton width for the assigned power, and $w_\mathrm{in}$ is the incident beam waist. As in the potential expressed by \eqref{eq:particle_potential}, the potential depends on the initial beam width \cite{Karimi:2016}.

\subsection{Initial condition for Gaussian beams}

The normalized wavefunction of a Gaussian beam can be written as $\psi = \left[2/(\pi w^2) \right]^{1/4} e^{-x^2/w^2} e^{i \sigma x^2}$, where $\sigma$ is related at the entrance facet to the curvature radius $R_\mathrm{in}= R_\mathrm{0}R_\mathrm{NL}/(R_\mathrm{0}+ R_\mathrm{NL})$ via $\sigma = k_0n_0/(2R_\mathrm{in})$. The direct computation of $\langle p^2 \rangle$ provides \cite{Sakurai:1994}
\begin{equation} \label{eq:psquare}
    \left\langle p^2 \right \rangle = \sigma^2 w^2 + \frac{1}{w^2}.
\end{equation}
This expression holds for widths $w$ and, consequently, curvature radii $R$ dependent in any manner on $z$. For Gaussian beams in a homogeneous medium, we retrieve the correct form $ \left\langle p^2 \right \rangle = 1/w_0^2$, which is, in fact, independent of the propagation distance $z$. When applied to the entrance facet of a nonlinear material, Eq.~\eqref{eq:psquare} provides $\langle p^2 \rangle_\mathrm{in} = [k_0 n_0/(2R_\mathrm{in})]^2 w_\mathrm{in}^2 + 1/w_\mathrm{in}^2$, that is, linear and nonlinear curvature radii are intrinsically intertwined in determining the initial condition. \\
We can now relate the initial quantity $\langle p^2 \rangle_\mathrm{in}$ with the boundary condition Eq.~\eqref{eq:boundary_R} to impose on the radius of curvature.
Using Eq.~\eqref{eq:ansatz_a} to compute $R_\mathrm{NL}$, the expression for $B$ provided by the RHS of Eq.~\eqref{eq:master_equation_with_boundary} % Eq.~\eqref{eq:bc_der2}
can be recast as follows
%\begin{align}
%   B &= \frac{2 w^2_\mathrm{in}}{R^2_\mathrm{NL}} + \frac{4 w^2_\mathrm{in}}{R_\mathrm{in}R_\mathrm{NL}} - \left( \frac{2}{\pi}\right)^N \frac{2  \sigma_n n_{2N} P^N}{n_0 w_\mathrm{in}^{2M}} \nonumber \\
%    & + \frac{8}{k_0^2 n_0^2 w_0^2} . \label{eq:B_gaussian_beams}
%\end{align}
%\begin{align}
%   B &= \frac{2 \gamma^2 P^{2N}}{k_0^2 n_0^2 w^{4M-2}_\mathrm{in}} + \frac{4 \gamma P^N}{k_0 n_0 R_\mathrm{in}  w^{2(M-1)}_\mathrm{in}} -  \frac{2  \gamma P^N}{n_0 w_\mathrm{in}^{2(M-1)}} \nonumber \\
 %   & + \frac{8}{k_0^2 n_0^2 w_0^2} . \label{eq:B_gaussian_beams}
%\end{align}
\begin{align}
   B &= \frac{2 \gamma^2 P^{2N}}{k_0^2 n_0^2 w^{4M-2}_\mathrm{in}} + \frac{4 \gamma P^N}{k_0 n_0 R_\mathrm{0}  w^{2(M-1)}_\mathrm{in}} -  \frac{2  \gamma P^N}{n_0 (M-1) w_\mathrm{in}^{2(M-1)}} \nonumber \\
    & + \frac{8}{k_0^2 n_0^2 w_0^2} . \label{eq:B_gaussian_beams}
\end{align}
In Eq.~\eqref{eq:B_gaussian_beams}, the first line provides the nonlinear contribution to the convexity of $w^2$, whereas the second line provides the linear contribution, the latter dependent only on the beam waist $w_0$, as it should be. It is easy to prove that the first term in Eq.~\eqref{eq:B_gaussian_beams} is negligible whenever the nonlinear effects are perturbative, that is, $|V_\mathrm{NL}|\ll k_0n_0$.

\section{QHO form for different nonlinearities}
\label{sec:QHO}
The integer $N$ is determined by the nonlinear mechanism, as well. In this section we will discuss some examples for the most common types of nonlinearities.

\subsection{Local nonlinearities}

Let us now compute the coefficient $a$ for different types of local nonlinearities. Taking into account Eq.~\eqref{eq:VNL}, we first set $V_\mathrm{NL}=-k_0 n_\mathrm{2N}I^N$, with $n_\mathrm{2N}$ thus being a sort of generalized Kerr coefficient and $N\in \mathbb{N}$. %Let us start from the simpler Kerr effect.
If we directly compute the derivative of the nonlinear index well, the parabolic coefficient will not be accurate given that the nonlinear perturbations, proportional to $I^N$, are narrower than the intensity profile. We thus need to define an effective parabolic coefficient that provides a good estimation of the force of nonlinear origin acting on the beam width $w$.
In analogy with the role of curvature radius in diffractive spreading, we define the focusing strength $F$ as the average over the beam cross-section of the Laplacian of $V_\mathrm{NL}$, that is, $F=\iint{I \nabla^2 V_\mathrm{NL} dxdy}/\iint{Idxdy}$. We now match $F$ computed by assuming the parabolic approximation \eqref{eq:parabolic_approximation} and the full form given by $V_\mathrm{NL}$. For radially symmetric Gaussian transverse profiles $I=\frac{2P}{ \pi w^2 } e^{\frac{-2 r^2}{w^2}} $, the effective coefficient $a_\mathrm{eff}$ reads
%\begin{align}
%    a_\mathrm{eff} &=   \frac{2\sqrt{2} n_\mathrm{2M}}{ \sqrt{\pi(M+1)}w} \left(\frac{2P}{\pi w^2} \right)^M {\int_{-\infty}^\infty{e^{-2\frac{x^2}{w^2}} \frac{d^2 e^{-2M\frac{x^2}{w^2}}}{dx^2}dx}} %{\int_{-\infty}^\infty{e^{-2x^2/w^2}dx}} 
 %   \nonumber \\ &= \sigma_M  n_\mathrm{2M} \left(\frac{2P}{\pi } \right)^M \frac{1}{w^{2(M+1)}} . \label{eq:a_eff}
%\end{align}
\begin{align}
    a_\mathrm{eff} &=   \frac{2\sqrt{2} n_\mathrm{2N}}{ \sqrt{\pi(N+1)}w} \left(\frac{2P}{\pi w^2} \right)^N {\int_{-\infty}^\infty{e^{-2\frac{x^2}{w^2}} \frac{d^2 e^{-2N\frac{x^2}{w^2}}}{dx^2}dx}} %{\int_{-\infty}^\infty{e^{-2x^2/w^2}dx}} 
    \nonumber \\ &= \sigma_N  n_\mathrm{2N} \left(\frac{2P}{\pi } \right)^N \frac{1}{w^{2(N+1)}} . \label{eq:a_eff}
\end{align}
The Gaussian ansatz for $I$ is coherent with the parabolic approximation for the nonlinear potential \cite{Vanicek:2023}. \\
When comparing with Eq.~\eqref{eq:ansatz_a}, we find $M=N+1$; %the dependence on the power $P$ requires $N=M$;
for local nonlinearities $\gamma$ is  
\begin{equation} \label{eq:gamma}
    \gamma(N) =  \left(\frac{2}{\pi}\right)^{N}  {\sigma_{N} n_\mathrm{2N} }.
\end{equation}
%\begin{equation}
    %\gamma(w;N) =  \left(\frac{2}{\pi}\right)^{N}  \frac{\sigma_{N} n_\mathrm{2N} }{w^2}.
%\end{equation}
%Comparison with Eq.~\eqref{eq:ansatz_a} yields $N=M+1$ and $\kappa=  \sigma_M n_\mathrm{2M} (2P/\pi)^{N-1}$, in turn providing for $\gamma$
%\begin{equation}
%    \gamma_N =  \left(\frac{2}{\pi}\right)^{N-1}  \frac{\sigma_{N-1} n_\mathrm{2(N-1)} }{w^2}.
%\end{equation}
%As stated when introducing the ansatz \eqref{eq:ansatz_a}, the averaging over the wave cross-section results in a coefficient $\gamma$ depending on the width of the wave itself. \\
The fit coefficient $\sigma_N$ is found by solving the  integral in Eq.~\eqref{eq:a_eff}
\begin{equation}
    \sigma_N = \frac{4N}{\left(N+1 \right)^{{2}}}.
\end{equation}
For the local Kerr effect in (2+1)D geometries $V_\mathrm{NL}= -k_0 n_2 I$, it is then $N=1$, $M=2$, and %$\sigma_2=4/(3\sqrt{3})\approx 0.77$.
$\sigma_1=1$.

\subsection{Nonlocal nonlinearities}

Nonlocal nonlinear effects often arise from some kind of diffusion whose source is the optical intensity \cite{Suter:1993}. These phenomena can often be modeled using a screened Poisson equation
\begin{equation} \label{eq:diffusion}
    \nabla^2 V_\mathrm{NL} - \left(\frac{\pi}{l}\right)^2 V_\mathrm{NL} = \frac{n_\mathrm{NL} I}{k_0} ,
\end{equation}
where the nonlocal character of the nonlinear response is fixed by $l$. Focusing condition requires $n_\mathrm{NL}>0$, resulting indeed in a negative potential $V_\mathrm{NL}$. When the size of the cell is smaller than the screening length $l$, Eq.~\eqref{eq:diffusion} turns into a Poisson equation, featuring an amount of nonlocality given by the minimum size of the cell \cite{Rothschild:2006}. In the opposite limit, for $l\rightarrow 0$ the solution of Eq.~\eqref{eq:diffusion} is
\begin{equation} \label{eq:VNL_local_limit}
    V_\mathrm{NL} = - \frac{n_\mathrm{NL}}{k_0}\left(\frac{l}{\pi} \right)^2 I
\end{equation}
that is, the material behaves like a local Kerr medium encompassing a Kerr parameter $n_2 = n_\mathrm{NL}l^2/\pi^2$. \\ 
%In the accessible soliton limit, the nonlinear index well is much broader than the beam itself, allowing the approximation of the full nonlinear potential with a parabolic profile of strength. 
Assuming cylindrical symmetry and sampling Eq.~\eqref{eq:diffusion} around the symmetry axis $(x=0,y=0)$, we find \cite{Conti:2004} 
\begin{equation} \label{eq:a_SM}
    a =   \frac{n_\mathrm{NL} P}{\pi w^2} + \left(\frac{\pi}{l}\right)^2 \frac{V_0}{2 }.
\end{equation}
We notice that $V_0$ is negative for focusing nonlinearities, thus a finite $l$ is decreasing the strength of the QHO. \\
 To rewrite the general nonlocal case in the form given by Eq.~\eqref{eq:ansatz_a}, we need to express $V_0$ as a power series with respect to the beam width $w$. In Appendix~\ref{sec:V0_expansion} it is demonstrated that $V_0 = n_\mathrm{NL} P \sum_{j=-1}^\infty c_j w^j$. Substituting back into Eq.~\eqref{eq:a_SM} yields
 \begin{equation} \label{eq:a_nonlocal_general}
     a = n_\mathrm{NL} P \left(  \frac{1}{\pi w^2} +  \left(\frac{\pi}{l}\right)^2 \sum_{j=-1}^\infty c_j (w^2)^{\frac{j}{2}} \right).
 \end{equation} 
Thus, unlike the local nonlinearities discussed above, the nonlinear potential in the nonlocal case contains terms of the type established by Eq.~\eqref{eq:particle_potential} encompassing $M=-j/2$, with $j =-1,0,1,2\ldots$.
In the highly nonlocal (HNL) case $l\rightarrow \infty$,
 Eq.~\eqref{eq:a_nonlocal_general} tells us that the HNL limit corresponds to $M=1$, thus corresponding to Eq.~\eqref{eq:particle_potential_HNL} for the fictitious potential $U$. When nonlocality is finite, the additional terms in $U$ weaken the self-confinement effect, finally yielding unstable solutions in the local limit $l/w\rightarrow 0$ corresponding to the Townes soliton. Although the HNL limit has also been applied to investigate shock waves in defocusing media \cite{Gentilini:2015}, in this Paper we will restrain our discussion to focusing nonlocal nonlinearity.

\section{Applications in different nonlinear media}
\label{sec:applications}
\subsection{Gaussian beams in a Kerr material}

We start by applying our method to the most common case in nonlinear optics, that is, Gaussian beams in a pure Kerr material. In Sec.~\ref{sec:NLGB_der2vanishing} we survey the properties of such beams stemming from the conservation of the second derivative of the second order moment. In Sec.~\ref{sec:NLGB_our_method} we frame such general properties in the context of our model. In Sec.~\ref{sec:NLGB_interpretation} we provide a physical interpretation for the behavior of the beam versus its input parameters, eventually comparing our theoretical results with numerical simulations.

\subsubsection{Gaussian beams in the nonlinear regime: general properties}
\label{sec:NLGB_der2vanishing}
\begin{figure}[h]
    \centering
  \includegraphics[width=0.99\linewidth]{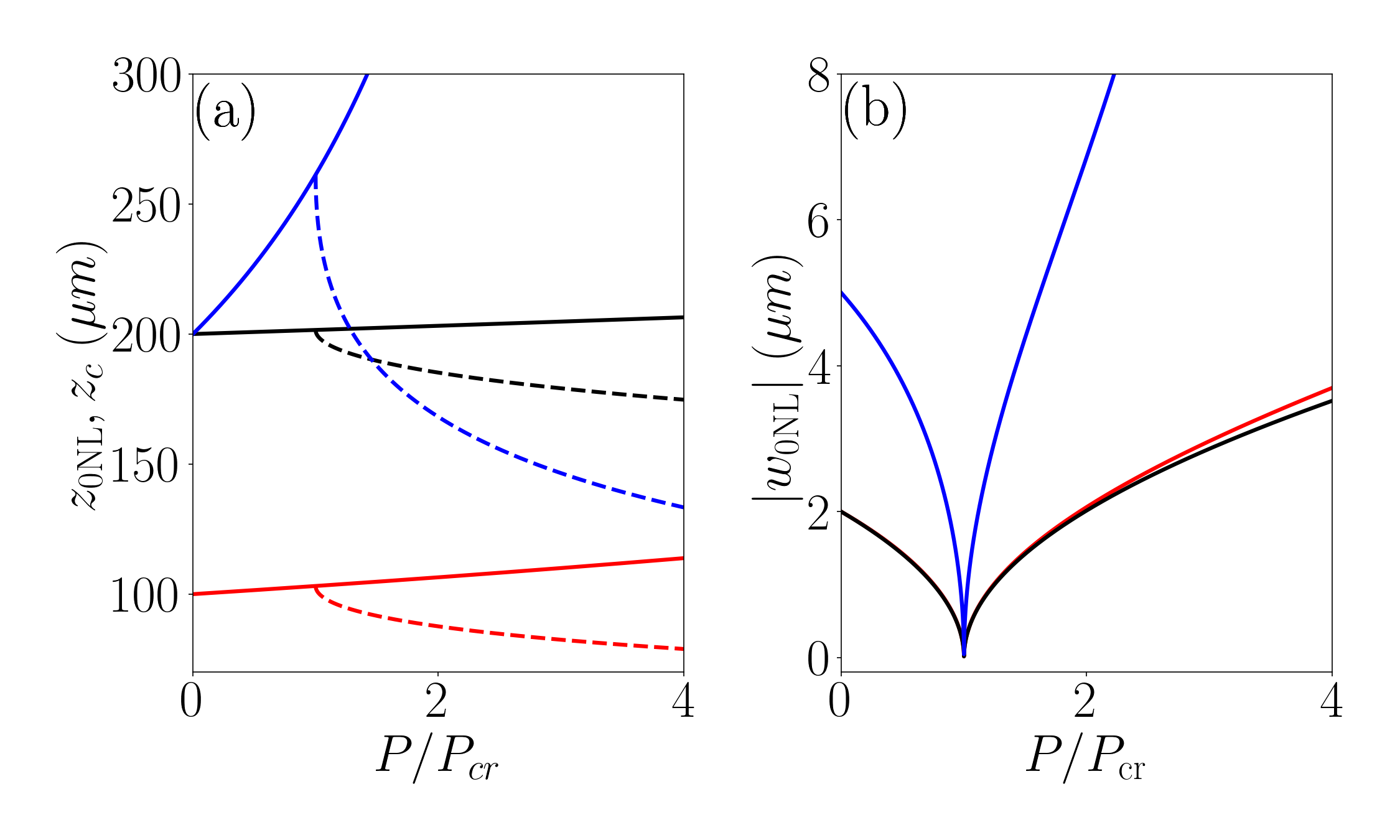}
    \caption{ Parameters of a NLGB versus the normalized power $P/P_\mathrm{cr}$ for a wavelength $\lambda=1064~$nm. (a) The nonlinear focal position $z_\mathrm{0NL}$ [solid lines, see Eq.~\eqref{eq:z0NL}] and the corresponding position of the catastrophic collapse $z_c$ [dashed lines, see Eq.~\eqref{eq:zc}]. (b) Absolute value of the nonlinear waist $w_\mathrm{0NL}$ as computed from Eq.~\eqref{eq:w0NL}. Red and black lines correspond to $w_0=2~\mu$m with linear focal position placed in $z=100~\mu$m and $200~\mu$m, respectively. Blue lines are computed for $w_0=5~\mu$m and linear focus placed in  $z=200~\mu$m. }
    \label{fig:NLGB_parameters}
\end{figure}

Nonlinear Gaussian beams (NLGBs) in a Kerr material have been widely investigated in literature \cite{Belanger:1983,Desaix:1991,Fibich:2000,Fan:2019}, including their relevance in ensuring mode-locking regime in ultrafast laser cavities \cite{Salin:1991,Brabec:1992} and their prominent role in laser machining \cite{Soileau:1989,Gattass:2008}. The aim of this section is to verify the validity of our theoretical approach to what is probably the most relevant case in optics. \\ 
According to Eq.~\eqref{eq:master_equation_with_boundary}, the curvature $d^2w^2/dz^2$ is conserved even in the nonlinear case, in accordance with the moment method \cite{Vlasov:1971}. Similarly to the linear case, we can therefore set \cite{Belanger:1983}
\begin{equation} \label{eq:NLGB}
w^2 =w^2_\mathrm{0NL}\left[ 1 + \alpha \left(z-z_\mathrm{0NL} \right)^2 \right] ,
\end{equation}
in turn yielding $d^2w^2/dz^2=2\alpha w^2_\mathrm{0NL}$, where the parameters of the NLGB $ w_\mathrm{0NL}$ and $z_\mathrm{0NL}$ depend on the input power $P$. Plugging a Gaussian solution into Eq.~\eqref{eq:NLSE} yields the following expression for $\alpha$
\begin{equation} \label{eq:alpha}
    \alpha = \frac{4}{n_0^2 k_0^2 w^4_\mathrm{0NL}} \left( 1 - \frac{P}{P_\mathrm{cr}}\right), 
\end{equation}
where the critical power $P_\mathrm{cr}$ is 
\begin{equation} \label{eq:Pcr}
    P_\mathrm{cr} = \frac{\lambda^2}{2\pi \sigma_1 n_0 n_2}.
\end{equation}
Following Ref.~\cite{Fibich:2000} and defining the critical power as $P_\mathrm{cr}=\rho \lambda^2/(4\pi n_0n_2)$, our result for Gaussian profiles provides $\rho = 2$, corresponding to the theoretical upper bound derived from the Hamiltonian and around $5\%$ larger than the numerical value 1.8962. \\
The parameter $\alpha$ is the nonlinear generalization of the Rayleigh distance $L_R$. In fact, in the linear regime $P\ll P_\mathrm{cr}$ Eq.~\eqref{eq:alpha} provides $\alpha = L_R^{-2}$ in agreement with the dynamics of Gaussian beams. By direct substitution, the beam width $w$ obeys
\begin{equation}
    \frac{d^2w}{dz^2} = \frac{4}{k_0^2 n_0^2 w^3} \left( 1 - \frac{P}{P_\mathrm{cr}}\right)
\end{equation}
in agreement with standard variational approach \cite{Fan:2019}, moment of methods \cite{Vlasov:1971} and self-similar (aka aberration-less) solutions \cite{Desaix:1991}. \\
To find the other two parameters $ w_\mathrm{0NL}$ and $z_\mathrm{0NL}$ we need to consider the initial conditions $w(z=0)=w_\mathrm{in}$ and Eq.~\eqref{eq:initial_curvature_radius}. For $N=1$, Eq.~\eqref{eq:a_eff} provides $a_\mathrm{eff}= 2 \sigma_1 n_2 P/\left(\pi w^2\right) $ for the QHO strength.  After introducing for the sake of compactness the three auxiliary quantities $c_1=n_0k_0 \pi w^4_\mathrm{in}- 2 n_2 \sigma_1 P R_\mathrm{in}$, $c_2=\pi R_\mathrm{in}n_0 k_0 w^4_\mathrm{in}$, and $c_3= 4/(n_0^2 k_0^2)(1-P/P_\mathrm{cr})$, the beam waist and focal position of the NLGB read
\begin{align}
    w_\mathrm{0NL}(P) &= \frac{w_\mathrm{in}}{\sqrt{1+\frac{c_1^2}{c_2^2 c_3}w_\mathrm{in}^4}}, \label{eq:w0NL} \\
    z_\mathrm{0NL}(P) &= \frac{c_1}{c_2 c_3} \frac{w_\mathrm{in}^4}{1+\frac{c_1^2}{c_2^2 c_3}w_\mathrm{in}^4}.  \label{eq:z0NL}
\end{align}
The position $z_c$ where catastrophic collapse occurs can be found by setting $w=0$. From Eq.~\eqref{eq:NLGB} we derive $z_c=z_\mathrm{0NL}- \alpha^{-0.5}$, finally providing
\begin{equation} \label{eq:zc}
    z_c = z_\mathrm{0NL} - \frac{n_0 k_0 w^2_\mathrm{0NL}}{2\sqrt{\frac{P}{P_\mathrm{cr}}-1}},
\end{equation}
in agreement with theoretical and experimental literature \cite{Kelly:1965,Vlasov:1971,Desaix:1991,Liu:2005,Lushnikov:2013}. \\ % the geometrical optics approximation \cite{Kelly:1965} and the method of moments~\cite{Vlasov:1971}. \\
The behavior of the NLGB parameters are plotted in Fig.~\ref{fig:NLGB_parameters} \cite{Hunter:2007}. The focal position $z_\mathrm{0NL}$ slightly changes with the power for $P<P_\mathrm{cr}$ when the initial beam is tightly focused: diffraction dominates over self-focusing. Once the catastrophic collapse kicks in (i.e., $P>P_\mathrm{cr}$), a rapid drop $\propto \left(1-P/P_\mathrm{cr}\right)^{-1/2}$ occurs. The shape of the drop is almost independent of the focal position $z_0$ (visually, a simple vertical shift connects black and dashed lines in Fig.~\ref{fig:NLGB_parameters}). Conversely, the wider the beam linear waist $w_0$ is, the more abrupt the decrease is. Finally, the nonlinear beam waist $w_\mathrm{0NL}$ is strictly positive and vanishes at $P=P_\mathrm{cr}$. 

\subsubsection{Correspondence with our method}
\label{sec:NLGB_our_method}
%In Eq.~\eqref{eq:master_equation_with_boundary} 
The integration constant $B$ provided by Eq.~\eqref{eq:B_gaussian_beams} takes the value
%\begin{equation} \label{eq:B_boundary_conditions}
%    B= \frac{8}{k_0^2 n_0^2 w_{0P}^2} - \frac{4 \sigma_1 n_2 P}{n_0 \pi w_\mathrm{in}^2}.
%\end{equation}
\begin{equation} \label{eq:B_boundary_conditions}
    B=  \frac{ 4\sigma_1 n_2 P}{ \pi n_0 w^2_\mathrm{in}} \left(\frac{2}{k_0 R_\mathrm{0}} -1 \right) + \frac{8}{k_0^2 n_0^2 w_{0}^2}.
\end{equation}
%We will start our discussion taking $w_\mathrm{0P}$ independent from the power and equal to the linear waist $w_0$: the validity of such assumption will be discussed at the end of the section. 
%For a fixed focal position $z_0$ and input width $w_\mathrm{in}$, $w_0$ is determined by the curvature radius $R_\mathrm{in}$. \\
In graphical terms, the potential $U$ for the effective particle is a straight line, whose slope depends on the interplay between diffraction and self-focusing. The condition $B>0$ is necessary for a Gaussian-like beam obeying Eq.~\eqref{eq:NLGB} to be a valid solution with a width $w$ following a convex curve; for $B<0$ the square width versus $z$ will become concave, that is, catastrophic collapse is ensured. \\
Equation~\eqref{eq:B_boundary_conditions} can then be recast as
\begin{equation} \label{eq:B_Pstar}
    B = \frac{8 }{n_0^2 k_0^2 w_{0}^2} \left(1- \chi \frac{w_0^2}{w^2_\mathrm{in}} \frac{P}{P_\mathrm{cr}} \right).
\end{equation}
with $\chi = k_0 R_\mathrm{0}/(k_0 R_\mathrm{0}-2)$. 
After rewriting Eq.~\eqref{eq:B_Pstar} in the form $B=8/(k_0^2n_0^2 w_0^2)(1-P/P^*_\mathrm{cr})$, the transition on the sign of concavity can therefore be associated with a second critical power $P^*_\mathrm{cr}$ given by
\begin{equation} \label{eq:Pcr_general_case}
    P^*_\mathrm{cr} = \frac{ \lambda^2 }{ 2\pi \sigma_1 n_0  n_2 \chi} \frac{w_\mathrm{in}^2}{w_{0}^2}= \left(1- \frac{2}{k_0 R_\mathrm{0}} \right) \frac{w_\mathrm{in}^2}{w_{0}^2} P_\mathrm{cr}.
\end{equation}
Let us now discuss the behavior of this second threshold with respect to the input conditions.
%Directly from the definition of $\chi$ and the properties of Gaussian beams, parameter $\chi$ is always larger than $k_0 L/(k_0L-1)\approx 1 + \frac{1}{k_0 L}$, where now the smallness parameter is proportional to $\lambda / w_0$. Thus, focusing parameter $\chi$ is always larger than 1; furthermore, $\chi$ substantially differs from unity only for highly nonparaxial beams where $w_0\sim \lambda$.
Before discussing the general behavior of $P^*_\mathrm{cr}$, we focus our attention on two limits: focus placed either on the interface or well inside the sample. When $R_\mathrm{0}\rightarrow\infty$ with $z_0=0$, the condition $B=0$ provides $P=P_\mathrm{cr}$, permitting to rewrite Eq.~\eqref{eq:B_Pstar} as $B= \frac{8}{k_0^2 n_0^2 w_0^2}(1-P/P_\mathrm{cr})$. When the input wavefront is not flat and focusing is tight ($z_0/L\ll 1$), we get $\chi \approx 1$; hence, the condition $B=0$ provides $P^*_\mathrm{cr}= (w^2_\mathrm{in}/w_0^2) P_\mathrm{cr} >P_\mathrm{cr}$.
In the general case, from the properties of Gaussian beams, we find $\text{max}(\left|2/(k_0 R_\mathrm{0} )\right|) = [1/(2\pi^2 n_0)] \left(\lambda /w_0 \right)^2$. The term between round brackets in Eq.~\eqref{eq:Pcr_general_case} is always very close to unity, with appreciable deviations occurring only for highly nonparaxial beams. Eventually, $P^*_\mathrm{cr}\approx (w^2_\mathrm{in}/w_0^2) P_\mathrm{cr} \geq P_\mathrm{cr}$ holds valid for every input condition accurately modeled by a scalar model for the optical propagation \cite{Lax:1975}.

%In Appendix~\ref{sec:NL_boundaries} it is also shown that $w_\mathrm{0P}\approx w_0$ is a very good approximation when $P\le P^*_\mathrm{cr}$. \\ 
%Equation~\eqref{eq:B_boundary_conditions} can then be recast as $B= \frac{8}{k_0^2 n_0^2 w_0^2}(1-P/P^*_\mathrm{cr})$.
%Eq.~\eqref{eq:Pcr} and Eq.~\eqref{eq:Pcr_general_case} match when $w_\mathrm{in}=w_{0P}\approx w_0$, that is, the incident beam shows a flat phase-front at the entrance of the nonlinear medium. 

%More generally, the definition of beam waist dictates $P^*_\mathrm{cr}\geq P_\mathrm{cr}$, its value being independent from the sign of the initial curvature radius $R_\mathrm{in}$.  %Thus, our approach suggests that the critical power for fundamental Gaussian beams depends on the launch condition, and it is not an intrinsic property of the material, as predicted by Eq.~\eqref{eq:Pcr}. A dependence from the focusing conditions for $P_\mathrm{cr}$ has been experimentally observed in Ref.~\cite{Polynkin:2013} in the case of ultrashort pulses, whereas its value for higher-order Gaussian beams has been addressed using the variational approach \cite{Fan:2019}. \\
Equating Eq.~\eqref{eq:B_Pstar} with the second derivative extracted from Eq.~\eqref{eq:NLGB}, we find the relationship between the waist $w_{0}$ and its  counterpart $w_\mathrm{0NL}$ accounting for the self-focusing effect 
\begin{equation} \label{eq:w0NL_vs_w0}
    w_\mathrm{0NL}^2 = \frac{1-\frac{P}{P_\mathrm{cr}}}{1-\frac{P}{P^*_\mathrm{cr}}} w^2_{0}.
\end{equation}
The square of the nonlinear waist vanishes when $P=P_\mathrm{cr}$ [see Fig.~\ref{fig:NLGB_parameters}(b)], turns negative for $P_\mathrm{cr}<P<P_\mathrm{cr}^*$, finally diverging when $P=P^*_\mathrm{cr}$ (in the latter case, $w^2(z) \propto  z$). \\
Equation~\eqref{eq:w0NL_vs_w0} paves the way to a more intuitive picture of how the waist $w_\mathrm{0NL}$ depends on the power $P$: for planar input wavefronts, $w_\mathrm{0NL}\approx w_0$ for all the input powers; for $P\ll P^*_\mathrm{cr}$ or for input widths $w_\mathrm{in}$ large enough to fulfill $w_0^2/w_\mathrm{in}^2 \ll P_\mathrm{cr}/P$, $w_\mathrm{0NL}^2\approx \left(1 - \frac{P}{P_\mathrm{cr}} \right)w^2_0$, that is, the nonlinear waist does not depend on the focal position in the linear regime $z_0$, in agreement with Fig.~\ref{fig:NLGB_parameters}(b). Other simplified formulae of interest for tight focusing conditions typically used in laser micro-machining \cite{Soileau:1989,Gattass:2008} are provided in Appendix~\ref{sec:tight_focusing}. \\
Finally, from Eq.~\eqref{eq:NLGB} and Eq.~\eqref{eq:alpha}  concavity $d^2w^2/dz^2$ can be written in the alternative form
\begin{equation}
B = \frac{8}{n_0^2 k_0^2 w^2_\mathrm{0NL}}\left(1- \frac{P}{P_\mathrm{cr}} \right),
\end{equation}
yielding Eq.~\eqref{eq:B_Pstar} once Eq.~\eqref{eq:w0NL_vs_w0} is plugged into: when crossing the critical power $P=P_\mathrm{cr}$, the sign of the concavity $B$ does not switch, although $w_\mathrm{0NL}$ is vanishing.

\begin{figure}[h]
    \centering
  \includegraphics[width=0.99\linewidth]{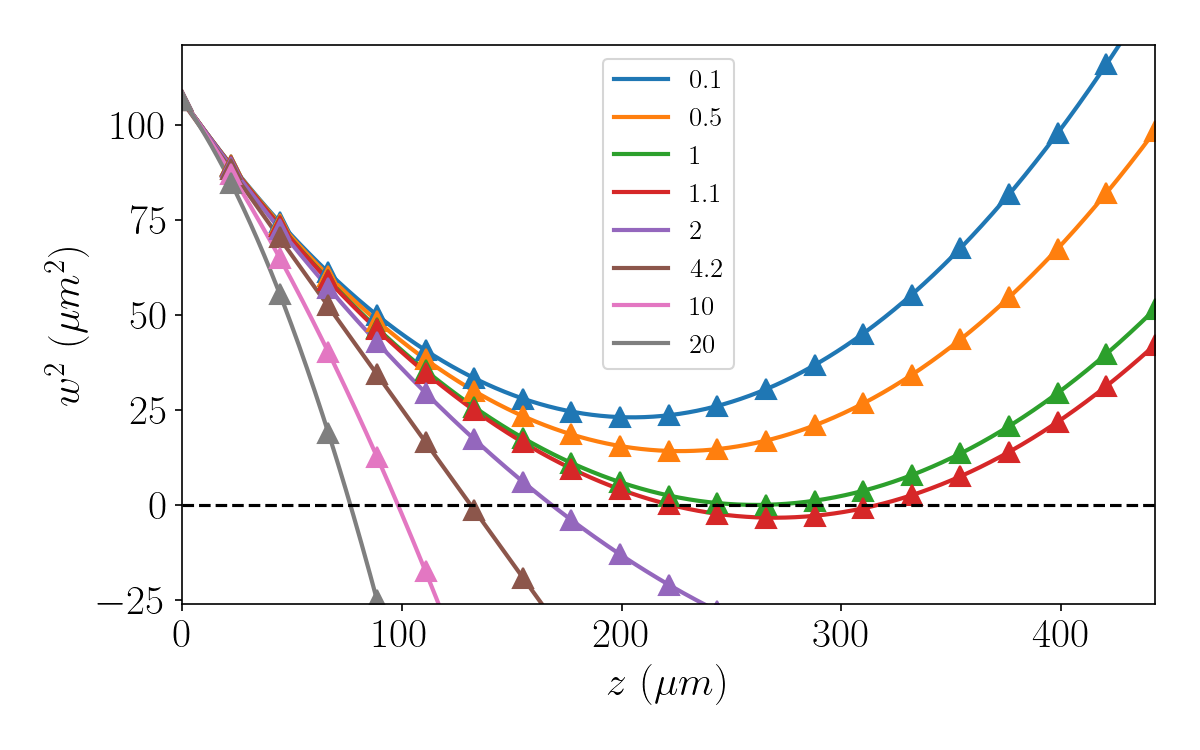}
    \caption{ Square width $w^2$ versus the propagation coordinate $z$ for different values $P/P_\mathrm{cr}$ as reported in the legend. Solid lines and triangles correspond to solutions from Eqs.~(\ref{eq:w0NL}-\ref{eq:z0NL}) and from Eq.~\eqref{eq:B_Pstar}, respectively. As direct consequence of the singularity at the critical power, the regions where $w^2<0$ have no physical meaning. Here $\lambda=1064~$nm, $w_0=5~\mu$m, and $z_0=200~\mu$m. }
    \label{fig:NLGB_width_vs_z}
\end{figure}

\subsubsection{Physical interpretation}

\label{sec:NLGB_interpretation}

We now wrap up all the previous mathematical results and provide a coherent physical interpretation of the wave dynamics predicted by our model, finally cross-checking its correctness with the available literature.  Given that $P^*_\mathrm{cr}\geq P_\mathrm{cr}$, the relevant threshold for catastrophic collapse is always $P_\mathrm{cr}$, regardless of the launch conditions. It is indeed well known that the critical power does not depend on the focusing conditions, but only on the transverse shape of the beam \cite{Vlasov:1971,Fibich:2000,Fan:2019}. If $P^*_\mathrm{cr}> P_\mathrm{cr}$, the collapse will occur when the second derivative $d^2 w^2/dz^2$ is still positive, but the nonlinear beam waist vanishes, see Eq.~\eqref{eq:w0NL_vs_w0}.  In the special case $w_\mathrm{in}=w_0$, the nonlinear waist is identical to the linear value, but the concavity of the curve changes sign according to the boundary condition provided by Eq.~\eqref{eq:B_boundary_conditions}. The exemplificative behavior of a NLGB featuring $z_0 \neq 0$ is shown in Fig.~\ref{fig:NLGB_width_vs_z}. For $P<P_\mathrm{cr}$, the beam follows the standard Gaussian profile with a nonlinear correction for the Rayleigh distance, as provided by Eq.~\eqref{eq:alpha}. At $P=P_\mathrm{cr}$, the waist is vanishing, that is, a catastrophic collapse occurs at $z_\mathrm{0NL}$ [see Fig.~\ref{fig:NLGB_parameters}(a)]. Incidentally, the model itself ceases to be valid before this threshold, given that the NLSE holds valid only for scalar waves in the paraxial limit \cite{Lax:1975,Feit:1988}. For $P_\mathrm{cr}<P<P_\mathrm{cr}^*$ ($P_\mathrm{cr}^*\approx4.2P_\mathrm{cr}$ in the plotted case), the concavity remains positive, with the position of the catastrophic collapse $z_c$ moving towards the entrance facet, see Fig.~\ref{fig:NLGB_parameters}(a). At $P=P^*_\mathrm{cr}$ the square width follows a straight line, eventually becoming a convex curve for further increases of the input power. The equivalent energy of the particle and its usage to explain the beam dynamics is provided in Appendix~\ref{sec:energy_kerr}.
Such a behavior agrees with the thin-lens transformation introduced by Talanov and showing that the behavior for finite $R_\mathrm{0}$ can be deduced from the flat wavefront case $w_\mathrm{in}=w_0$ \cite{Talanov:1970,Malkin:1993}.

\begin{figure*}[ht]
    \centering
  \includegraphics[width=0.99\linewidth]{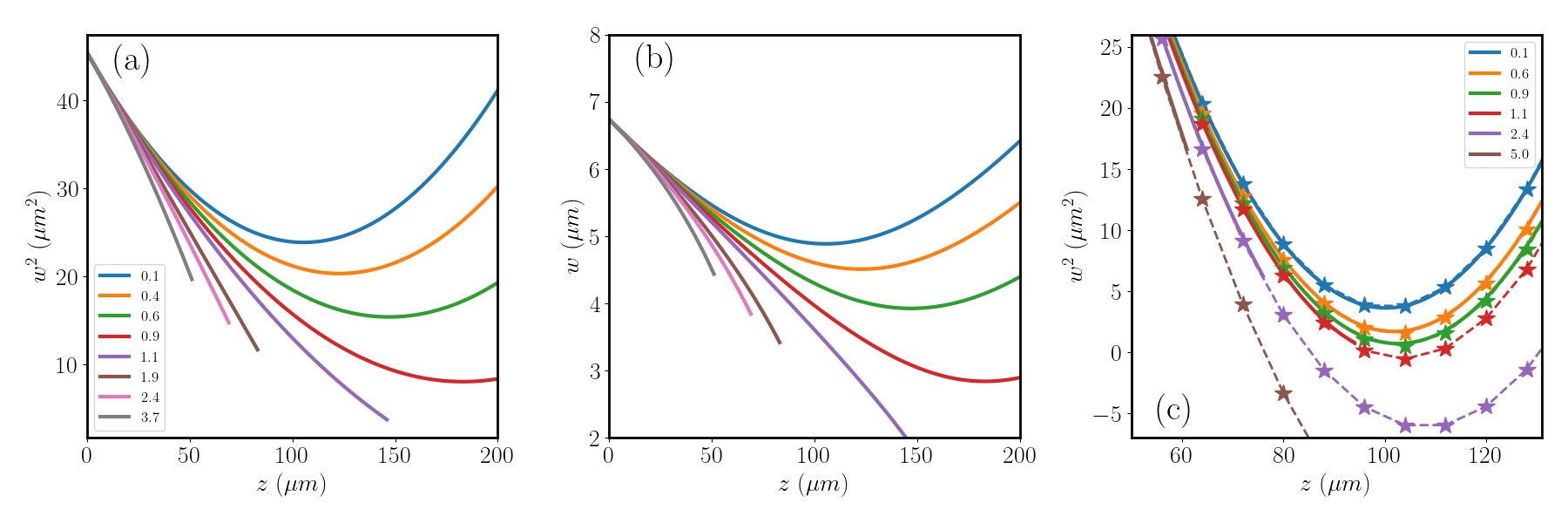}
    \caption{ Comparison between theory and simulations in the Kerr case. Numerically-computed square of the beam width $w^2$ (a) and the corresponding beam width $w$ (b)  versus the propagation distance $z$ for $w_0=5~\mu$m. (c) Squared width $w^2$ retrieved from numerical simulations (solid lines) and predicted theoretically (dashed lines with symbol) for $w_0=2~\mu$m. Legends provide the normalized power $P/P_\mathrm{cr}$, the latter including the same set of values in panel (a) and (b). In all the panels $\lambda=1064~$nm, $n_0=1.5$, and $z_0=100~\mu$m. }
    \label{fig:comparison_Kerr}
\end{figure*}

We verified our results performing numerical simulations of the NLSE, see Appendix~\ref{sec:numerics} for details on the employed method. The comparison between theory and simulations is shown in Fig.~\ref{fig:comparison_Kerr} for the case when $z_0=100~\mu$m and waist of 5~$\mu$m (a-b) and 2~$\mu$m (c). Numerical simulations are halted at the propagation distances where the field peak reaches $1\times 10^{4}$ the peak at the entrance facet: indeed, the numerical error mostly impacts the computation of the phase profile, making the simulations unreliable when the nonlinear focusing is too abrupt. Due to the different focal position $z_0$ with respect to Fig.~\ref{fig:NLGB_width_vs_z}, for a waist $w_0$ of 5~$\mu$m it is now $P^*_\mathrm{cr}\approx 1.8 P_\mathrm{cr}$: accordingly, the change in convexity in Fig.~\ref{fig:comparison_Kerr}(a) occurs between the violet and the brown curve. As plotted in Fig.~\ref{fig:comparison_Kerr}(b), the change in the convexity of the curve $w$ versus $z$ takes place at a different input power given that $\frac{d^2 w^2}{dz^2}=2w\frac{d^2 w}{dz^2}+2\left(\frac{dw}{dz} \right)^2$.
A direct comparison between simulations and theoretical predictions from Eq.~\eqref{eq:NLGB} is provided in Fig.~\ref{fig:comparison_Kerr}(c). %showing a very good agreement between numerical simulations and theoretical prediction from Eq.~\eqref{eq:NLGB}.  
The two approaches are in very good agreement, confirming the validity of our method in the case of a purely Kerr local nonlinearity, even in the case of tight focusing. Interestingly from the point of view of material processing \cite{Gattass:2008}, simulations confirm that nonlinear changes in the intensity profile are minimized when the beam is strongly focused, thus confirming the findings shown in Fig.~\ref{fig:NLGB_parameters}.
%\begin{figure*}[t]
%    \centering
 % \includegraphics[width=0.95\linewidth]{theory_interplay.png}
  %  \caption{ Theoretical results for cubic-quintic media  when $\lambda=1064~$nm, $N_0=2$, and $n_2=5\times 10^{-18}$m$^{2}$W$^{-1}$. (a-b) Effective potential $U$ versus the square width $w^2$ for $n_4= - n_2\times 10^{-17}$m$^2$W$^{-1}$, for a fixed power $P=1.5P_\mathrm{cr}$ (a) or for a fixed waist $w_0=10~\mu$m (b). (c) Extreme width $w_\mathrm{ext}$ versus the normalized power $P/P_\mathrm{cr}$. Average width $w_\mathrm{av}$ (d), breathing period $\Lambda$ (e) and breathing amplitude $\Delta w$ (f) versus the normalized power $P/P_\mathrm{cr}$. In (d-f) solid and dashed lines correspond to $n_4 = n_2\times 10^{-17}$m$^2$W$^{-1}$ and $n_4 = n_2\times 10^{-16}$m$^2$W$^{-1}$, respectively. In panel (e) a period $\Lambda_0=1~$mm is taken as reference for the logarithmic scale.} 
    %\label{fig:potential_focus_defocus}
%\end{figure*}

\subsection{Interplay between focusing and defocusing nonlinearities}

In this subsection we consider the simultaneous presence of the Kerr effect with $n_2>0$ plus a higher-order nonlinearity featuring $N=N_0>1$, $V_\mathrm{NL}= -k_0\left(n_2 I + n_{2N_0}I^{N_0}\right)$ \cite{Wright:1995,Malkin:1993,Abdullaev:2001,Zhan:2002,Boudebs:2003,Falcao:2013}. %With respect to the integration constant, we can safely assume is determined by the Kerr lensing: at the entrance facet the beam won't be extremely narrow, making the contribution $\propto w^{-2}$ dominant over the one $\propto w^{-2N_0}$. 
The potential Eq.~\eqref{eq:particle_potential} reads
\begin{equation} \label{eq:potential_focus_defocus}
   U(w^2) =  \frac{-2\sigma_{N_0} n_{2N_0} }{n_0 N_0} \left(\frac{2P}{\pi}\right)^{N_0} \frac{1}{w^{2(N_0-1)}} - Bw^2,
\end{equation}
where the expression for $B$ is found from the boundary condition expressed by Eq.~\eqref{eq:B_gaussian_beams} 
\begin{align} \nonumber
    B &= \frac{4\gamma(1) P}{k_0 n_0 R_0 w_\mathrm{in}^2} \left( 1+  \frac{\gamma (N_0)}{\gamma (1)}\frac{P^{N_0-1}}{w_\mathrm{in}^{2(N_0-1)}} \right)  \\
    & - \frac{2\gamma(1) P}{n_0 w_\mathrm{in}^2} \left( 1+  \frac{1}{N_0}\frac{\gamma (N_0)}{ \gamma (1)}\frac{P^{N_0-1}}{w_\mathrm{in}^{2(N_0-1)}} \right) \nonumber \\
    &
    + \frac{8}{k_0^2 n_0^2 w_0^2} \nonumber %\label{eq:B_interplay_middle}
\end{align}
where the form of $\gamma$ is given by Eq.~\eqref{eq:gamma}. Recasting in terms of critical power we finally find
%Eq.~\eqref{eq:bc_der2}
%\begin{align} \nonumber
 %   B &= \frac{8}{k_0^2 n_0^2 w_0^2} \left[1 - \left(1+ \eta I_0^{N_0-1}  % \left(\frac{2P}{\pi w_\mathrm{0}^2} \right)^{N_0-1} 
  %  \right) \frac{P}{P^*_\mathrm{cr}} \right]  \\
   % & + \frac{4}{k_0 n_0 R_\mathrm{in}} \frac{2P}{\pi w^2_\mathrm{in}} \left( 1+ \eta I_0^{N_0-1} \right) \label{eq:B_interplay}
%\end{align}
\begin{align} \nonumber
    B &= \frac{8}{k_0^2 n_0^2 w_0^2} \left[1 - \left(1+ \frac{\eta I_0^{N_0-1}}{N_0}  % \left(\frac{2P}{\pi w_\mathrm{0}^2} \right)^{N_0-1} 
    \right) \frac{P}{P_\mathrm{cr}} \right]  \\
    & + \frac{4}{k_0 n_0 R_\mathrm{0}} \frac{2P}{\pi w^2_\mathrm{in}} \left( 1+ \eta I_0^{N_0-1} \right) \label{eq:B_interplay}
\end{align}
where $\eta = \frac{\sigma_{N_0} n_{2N_0}}{\sigma_1 n_2}= \frac{\sigma_{N_0}}{\sigma_1}\eta^\prime$ is the relative strength of the higher-order nonlinearity with respect to the Kerr effect, and $I_0=2P/(\pi w^2_0)$ is the maximum intensity at the input section. Eq.~\eqref{eq:B_interplay} is identical to the pure Kerr case provided by Eq.~\eqref{eq:B_Pstar} whenever the self-focusing Kerr potential dominates over the higher-order term at the interface, that is, $|\eta| I_0^{N_0-1}\ll 1$. Such a condition therefore depends on the peak intensity at the entrance $I_0$ and on the ratio between the two nonlinear coefficients $n_2$ and $n_{2N_0}$. % is fulfilled by standard nonlinear materials, at least for not too large powers.  

 If $n_{2N_0}>0$, the higher order nonlinearity is adding up in increasing the self-focusing, making the collapse even faster. If $n_{2N_0}<0$, the higher order nonlinearity is defocusing, thus counteracting the Kerr self-focusing effect. We will restrict our consideration to negative $n_{2N_0}$, that is, self-trapping is allowed. It is then evident from the potential Eq.~\eqref{eq:potential_focus_defocus} that a necessary condition for self-trapping is $B<0$. \\
 Hereafter, we will solely consider the case $w_\mathrm{in}=w_0$; in Eq.~\eqref{eq:B_interplay} the second term is therefore vanishing. The condition $B<0$ turns into
 \begin{equation} \label{eq:pcr_interplay}
     \frac{P}{P_\mathrm{cr}} - 1 >  \frac{|\Xi|}{P_\mathrm{cr}}  P^{N_0},
 \end{equation}
 where $\Xi= \eta \left[2\left/\left(\pi w_\mathrm{0}^{2}\right)\right.\right]^{N_0-1}/N_0 $. The limit $\eta\rightarrow 0$ converges to the pure Kerr case, as should be. For any value of $N_0$, the model predicts self-trapping in a finite interval $P_\mathrm{inf}<P<P_\mathrm{sup}$, corresponding to the two intersection points between the curves defined by the left and right terms of Eq.~\eqref{eq:pcr_interplay}. The lower threshold $P_\mathrm{inf}$ is due to the effective increase in the amount of diffractive spreading, thus inducing an increase in the critical power for self-focusing. The upper threshold $P_\mathrm{sup}$ is caused by the overall dominance of self-defocusing over self-focusing, thus inhibiting self-confinement in terms of bell-shaped bright solitons \cite{Wright:1995}. For large enough $\Xi$, the two curves become tangent to each other, thus $P_\mathrm{inf}=P_\mathrm{sup}\equiv P_\mathrm{th}$: further increases in $\Xi$ will then inhibit the self-trapping in the form of bell-shaped bright solitons, no matter what the input power $P$ is. 
 
Once ascertained the existence condition, let us focus on the properties of the self-trapped beams.
The local minimum of the potential Eq.~\eqref{eq:potential_focus_defocus} directly provides the average width $w_\mathrm{av}$
\begin{equation} \label{eq:wav_theory}
    %w_\mathrm{av} = \left( \chi\frac{N_0-2}{N_0-1} \right)^{\frac{1}{2(N_0-1)}}  \left[\frac{1}{P^{N_0}} \left(1- \frac{P}{P_\mathrm{cr}} \right)\right]^{-\frac{1}{2(N_0-1)}}, 
    %w_\mathrm{av} = \left[ -\frac{2 \sigma_{N_0-1}  n_{2(N_0-1)}  }{n_0 B} \frac{N_0-2}{N_0-1} \left( \frac{2P}{\pi}\right)^{N_0-1} \right] ^{-\frac{1}{2(N_0-1)}}, 
    w_\mathrm{av} = \left[ \frac{2 \sigma_{N_0}  n_{2N_0}  }{n_0 B}  \left( \frac{2P}{\pi}\right)^{N_0} \right] ^{\frac{1}{2N_0}}, 
\end{equation}
where the previous expression holds for any value of $N_0$. When $w_\mathrm{av}=w_0$, a shape-invariant spatial soliton is excited (in practice, breathing amplitude will be minimized given that  Gaussian is not the exact solution). The center of the oscillations $w_\mathrm{av}$, beyond the trivial dependence on the input power $P$, depends on the waist of the initial beam as well. 

%Whereas $w_\mathrm{ext}$ is independent from $w_0$, 
 
 %From Eq.~\eqref{eq:B_interplay} and under the condition $w_\mathrm{in}=w_0$, we derive the condition $P>P_\mathrm{cr}$, regardless of how large $N_0$ becomes, in agreement with Ref.~\cite{Wright:1995}.

\subsubsection{Soliton existence in cubic-quintic media}

  To test the physical soundness of our model, we now consider the most common case of a cubic-quintic material, that is, $N_0=2$ \cite{Falcao:2013,Reyna:2017}. Then $\eta = 1.125 \eta^\prime$. Solving Eq.~\eqref{eq:pcr_interplay} is equivalent to specify the range of values for which a concave parabola is positively valued: if existing, the interval for physically-sounded solutions will always be finite. From Eq.~\eqref{eq:pcr_interplay} it is straightforward to get $P_\mathrm{inf}= \left(1-\sqrt{1-4 |\Xi| P_\mathrm{cr}} \right)\left/\left(2|\Xi|\right)\right.$ and $P_\mathrm{sup}= \left(1+\sqrt{1-4 |\Xi| P_\mathrm{cr}} \right)\left/(2|\Xi|)\right.$. When the defocusing effect is small ($|\eta|\ll 1$, in turn implying $|\Xi| \ll 1$) with respect to the Kerr effect, the two power thresholds are $P_\mathrm{inf} \approx P_\mathrm{cr}$ and $P_\mathrm{sup}\approx P_\Xi - P_\mathrm{cr}$, where $P_\Xi=|\Xi|^{-1}=9\pi w_0^2 n_2/(8|n_4|)$. In terms of critical power, $P_\Xi = 9w_0^2\lambda^2 /(16n_0 |n_4| P_\mathrm{cr})$. 
  On the other side, physically-meaningful solutions cease to exist for a defocusing nonlinearity large enough to achieve $ P_\Xi =4 P_\mathrm{cr} $, that is, $|n_4|>9 w_0^2\lambda^2/(64n_0 P_\mathrm{cr}^2)$: lower and upper thresholds then become identical with $P_\mathrm{inf}=P_\mathrm{sup} = 2 P_\mathrm{cr} $. \\
  \begin{figure}[t]
    \centering
  \includegraphics[width=0.95\linewidth]{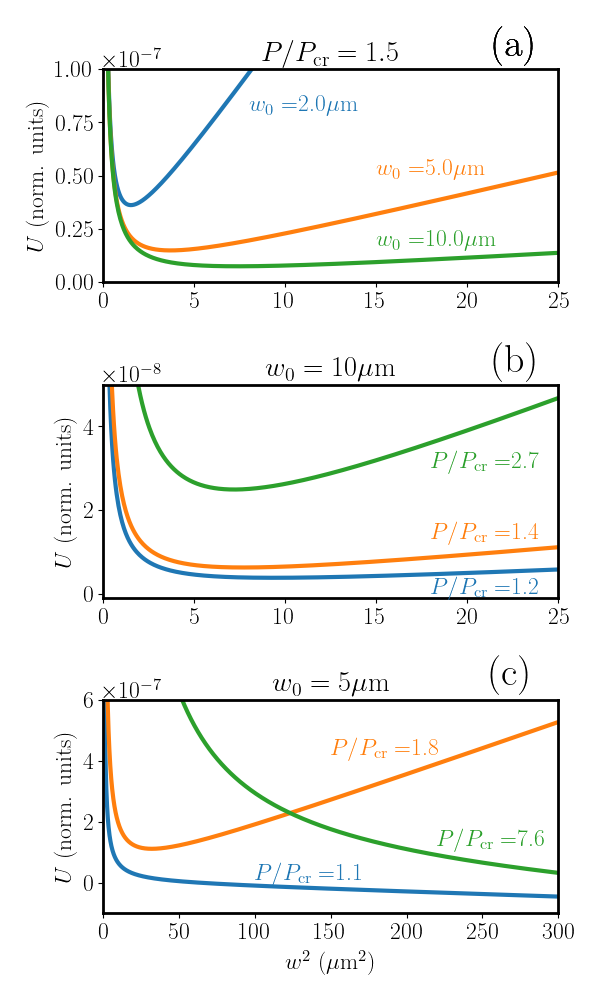}
    \caption{ Effective potential  versus the square width $w^2$ for cubic-quintic media  when $\lambda=1064~$nm, $N_0=2$, and $n_2=5\times 10^{-18}$m$^{2}$W$^{-1}$. (a-b) Effective potential for $|\eta^\prime| = |n_4|/n_2=1 \times 10^{-17}$m$^2$W$^{-1}$, for a fixed power $P=1.5P_\mathrm{cr}$ (a) or for a fixed waist $w_0=10~\mu$m (b). (c) Potential for three different powers when $|n_4|/n_2= 4.5 \times 10^{-16}$m$^2$W$^{-1}$; input width is now 5~$\mu$m.} 
    \label{fig:potential_focus_defocus}
\end{figure}
  %The threshold for the existence of bell-shaped self-confined beams is achieved for $|\Xi|=1/(4 P_\mathrm{cr})$, in turn providing $P_\mathrm{th}=2P_\mathrm{cr}$: 
  %As a direct consequence, $P=2P_\mathrm{cr}$ is an upper bound for powers supporting bell-shaped solitons, which surprisingly does not depend on the ratio $\eta$ between the two nonlinearities. 
  The condition for bell-shaped self-confinement $P_\mathrm{cr}<P_\Xi/4$ determines an upper limit the maximum degree of spatial localization achievable in a cubic-quintic material. Indeed, isolating the beam waist in the definition of $P_\Xi$, we can easily compute that
  %determines via $\Xi$ the minimum nonlinear waist $w_\mathrm{0NL}$ achievable in a given material for a fixed wavelength 
  %\begin{equation} \label{eq:min_wsol}
  %    w^\mathrm{min}_\mathrm{sol} = \sqrt{\frac{4\lambda^2\sigma_2 |n_4|}{\pi^2 n_0 \sigma_1^2 n_2^2}},
  %\end{equation}
  \begin{equation} \label{eq:min_wsol}
      w^\mathrm{min}_\mathrm{sol} = \sqrt{\frac{16\lambda^2 |n_4|}{ 9\pi^2 n_0 n_2^2}},
  \end{equation}
  which is indeed independent of the input power $P$. 
From now on, we will fix the Kerr coefficient at a given value, $n_2=5\times 10^{-18}$m$^{2}$W$^{-1}$, whereas the interplay between the two competing nonlinearities is modulated by varying $\eta^\prime$. 

  Let us discuss graphically the main properties of self-trapping. Figure~\ref{fig:potential_focus_defocus} illustrates the potential behavior versus $w^2$ for a fixed value of $\eta$ and different values for the initial waist $w_0$ and the input power $P$.
 In Fig.~\ref{fig:potential_focus_defocus}(a) the potential  for a fixed power greater than $P_\mathrm{inf}$ and three different input waists is shown. A local minimum, whose position provides the average width $w_\mathrm{av}$, occurs for all three values of $w_0$, that is, the condition $B<0$ is fulfilled in all three cases. The minimum position shifts towards the right for broader input beams, that is, stronger confinement is achieved for narrower input beams. Owing to the change in the potential shape, a much more pronounced breathing with longer periods and longer oscillation amplitudes are expected for broader input beams. Due to the shape of the effective potential $U$, the breathing oscillations will be strongly anharmonic in the general case, moreover showing a strong asymmetry between positive and negative half-periods, similarly to what has been demonstrated in the HNL case \cite{Alberucci:2016_1}. In Fig.~\ref{fig:potential_focus_defocus}(b) the dependence on the input power for a fixed initial waist is shown. Due to the self-defocusing, $w_\mathrm{av}$ does not monotonically decrease as the input power is increased. 
 To illustrate the presence of the two thresholds $P_\mathrm{inf}$ and $P_\mathrm{sup}$, in Figure~\ref{fig:potential_focus_defocus}(c) the defocusing effect (quantified by $n_4$) is increased 45 times with respect to panel (b). For the same purpose, the width of the input beam $w_0$ is halved to increase $|\Xi|$ even further. In doing so, the weight of the defocusing nonlinearity becomes much more relevant on the computation of $B$. An input power $P=1.1P_\mathrm{cr}$ (blue curve) is now lower than $P_\mathrm{inf}$, thus inhibiting self-trapping. On the other side, a power $P=7.6P_\mathrm{cr}$ (green curve) is larger than $P_\mathrm{sup}$, once again preventing Gaussian-shaped self-trapped solutions. 

\subsubsection{Comparison with numerically-computed solitary waves}

Next, we investigate how accurate our theory is in the case of shape-preserving solitons, i.e., in the absence of breathing \cite{Kivshar:2003,Conti:2004}. Theoretically, the soliton existence curve in the plane soliton width $w_\mathrm{sol}$ versus the normalized input power $P/P_\mathrm{cr}$ can be computed from Eq.~\eqref{eq:wav_theory} providing the average beam width. Operationally, for a fixed power Eq.~\eqref{eq:wav_theory} is evaluated versus the input beam width $w_0$, i.e., for different values of $B$; $w_\mathrm{sol}$ is then found by solving the equation $w_\mathrm{av}(w_0,P)=w_0$, that is, imposing that the initial beam width coincides with the position of the minimum of the effective potential $U(w^2)$. In terms of the mechanical analogy, we have an invariant soliton when the effective energy equals the potential energy at the bottom of the well, that is, the effective kinetic energy is vanishing.
 
\begin{figure}[t]
    \centering
  \includegraphics[width=0.95\linewidth]{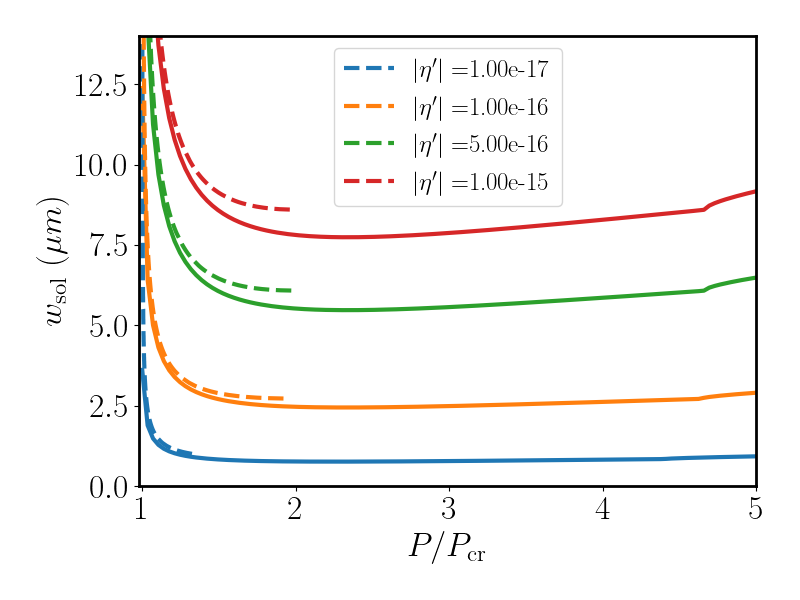}
    \caption{ Soliton width $w_\mathrm{sol}$ versus the normalized power $P/P_\mathrm{cr}$ computed by solving  numerically the nonlinear eigenvalue problem (solid lines) and predicted by Eq.~\eqref{eq:wav_theory} (dashed lines). Each color corresponds to a different ratio $\eta^\prime$ between the cubic and the quintic nonlinearities.} 
    \label{fig:comparison_eigenvalues}
\end{figure}

 Our theoretical results are compared with the numerical computation of the soliton profile in Fig.~\ref{fig:comparison_eigenvalues} where the soliton width $w_\mathrm{sol}$ versus the carried power $P/P_\mathrm{cr}$ is plotted, see Appendix~\ref{sec:NL_eigenvalue} for details on the numerical calculation. From first glance, good agreement is achieved whenever the theoretical solution exists. Quantitatively, numerical results follow the shape of the theoretical curve: for reference, theory overestimates the soliton width less than $10\%$ in the flat regions. The greatest discrepancy is the different existence interval: on one hand, in agreement with the discussion above, theory (dashed lines) predicts that a bell-shaped soliton exists only for $P<P_\mathrm{sup}$, no matter the value assumed by $\eta^\prime$;  % except that theory overestimates the soliton width between $20\%$ and $30\%$. 
 on the other hand, given there are no limitations on the shape of the transverse profile, the numerical simulations find a solution for any input power. \\
 The accuracy of the theory is quite good, also, when a deeper analysis is carried out. First, for $P\rightarrow P_\mathrm{cr}$ the lower threshold $P_\mathrm{inf}$ converges to $P_\mathrm{cr}$ (i.e., a soliton exists whenever the input power overcomes the critical value) regardless of $\eta$ because in the soliton case $w_0$ can increase at will, therefore making $P_\Xi$ arbitrarily large. In the opposite limit of large  powers, $P_\Xi$ gets smaller and smaller as the soliton narrows; when finally $P_\Xi=4P_\mathrm{cr}$, the soliton power cannot overcome $2P_\mathrm{cr}$. Accordingly, such as an upper threshold for the soliton power is easier to reach when defocusing is stronger.
 Nevertheless, in agreement with our theory, at $P=2P_\mathrm{cr}$ the beginning of the transition from bell-shaped to flat-top profiles is observed in the numerical solutions (see Fig.~\ref{fig:soliton_numerics} in the Appendix). The abrupt change in the numerically-computed $w_\mathrm{sol}$ just before $P=5P_\mathrm{cr}$ is indeed associated with the appearance of a dip in the center of the soliton, eventually developing into a ring-shaped profile for further increases in the input power \cite{Wright:1995}.

\subsubsection{Comparison with dynamical simulations}

\begin{figure}[t]
    \centering
  \includegraphics[width=0.95\linewidth]{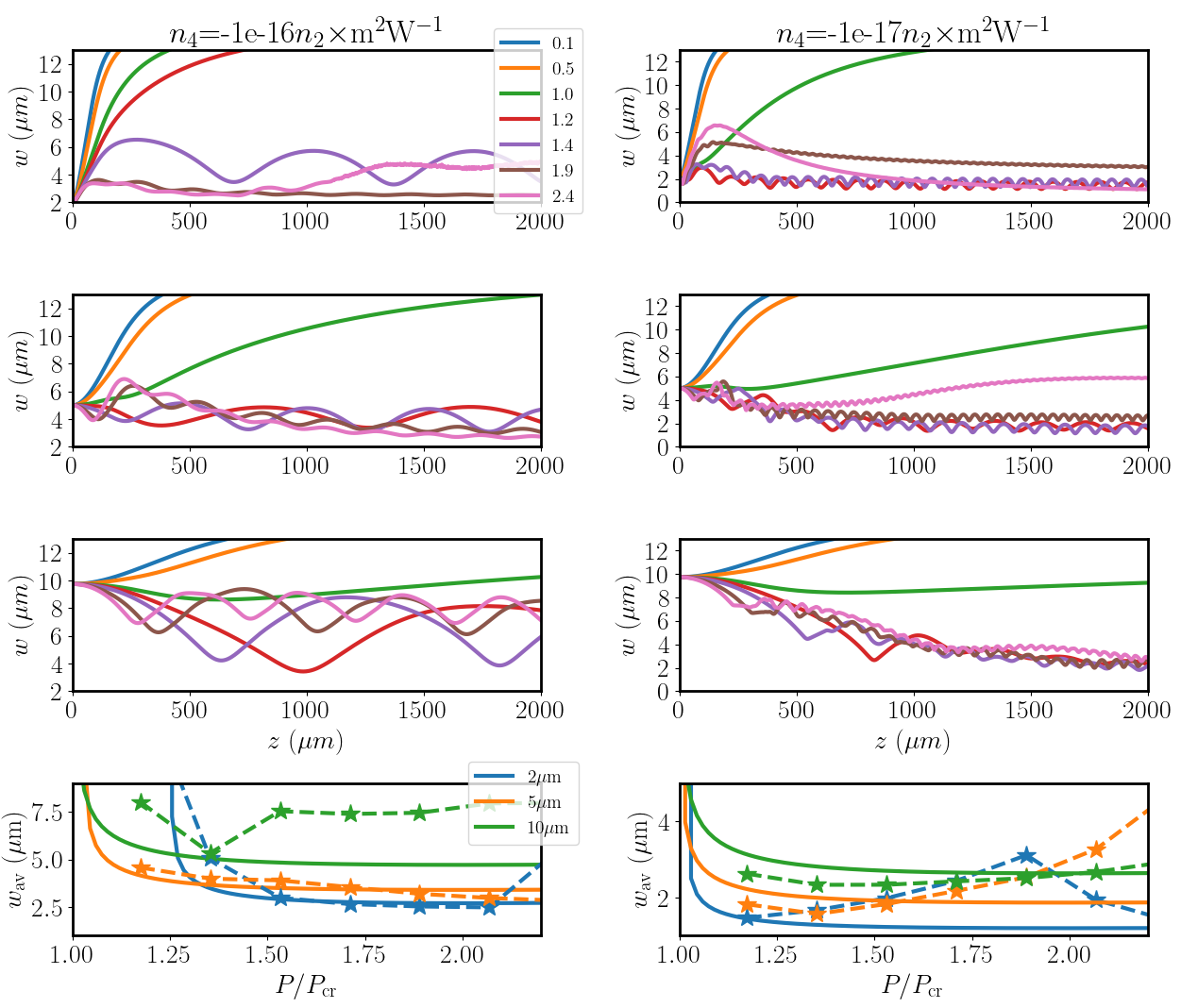}
    \caption{Top three rows: Beam width $w$ versus the propagation distance $z$ computed by the BPM simulations for the cubic-quintic nonlinearity. From top to bottom, rows correspond to an input beam waist $w_0$ of $2~\mu$m, $5~\mu$m, and $10~\mu$m, respectively. Each column corresponds to a different weight of the defocusing nonlinearity, as reported in the title. The legend in the first panel provides the color associated with each input power $P/P_\mathrm{cr}$. Bottom row: comparison between theoretical predictions from Eq.~\eqref{eq:wav_theory} (solid lines) and numerical simulations (dashed lines with stars). %Simulations parameters are the same of Fig.~\ref{fig:potential_focus_defocus}. 
    Here $\lambda=1064~$nm, $N_0=2$, and $n_2=5\times 10^{-18}$m$^{2}$W$^{-1}$.} 
    \label{fig:simulations_focus_defocus}
\end{figure}

In the next step we compare our theoretical predictions with full numerical simulations of the NLSE Eq.~\eqref{eq:NLSE}, see Appendix~\ref{sec:numerics} for a depiction of the code. Figure~\ref{fig:simulations_focus_defocus} shows the behavior in propagation of the beam width for three different input widths $w_0$ and two different ratios $\eta$. As expected and well known, self-trapping is dampened when the magnitude of the defocusing nonlinearity is increased (compare the two columns). In qualitative agreement with theory, all the self-trapped waves undergo an anharmonic breathing, whose features depend both on the input power and the initial input waist $w_0$. Simulations also confirm how the threshold for self-trapping, beyond the size of the defocusing nonlinearity, depends on the initial waist $w_0$. Quantitative comparison is shown in the bottom row of Fig.~\ref{fig:simulations_focus_defocus}. The numerical averaged beam width (symbols) is calculated in the interval $z\in [1.5\ 2]$mm, whereas the theoretical curves (solid lines) are computed from Eq.~\eqref{eq:wav_theory}. %Absence of the theoretical prediction curve means $B>0$, that is, no self-trapping. 
Whereas the order of magnitude and the relative dependence versus $w_0$ is the same with the two approaches, a much more accurate prediction is obtained when the input width is close to the condition for the excitation of a shape-preserving soliton: % width $w^\mathrm{min}_\mathrm{sol}$. 
 a good agreement for $w_\mathrm{av}$ versus the power is then reached for $|\eta |\geq 1\times 10^{-16}$m$^2$W$^{-1}$, that is, for defocusing nonlinearities strong enough to appreciable counteract the Kerr self-focusing already at powers near the critical value. \\
\begin{figure}[t]
    \centering
  \includegraphics[width=0.99\linewidth]{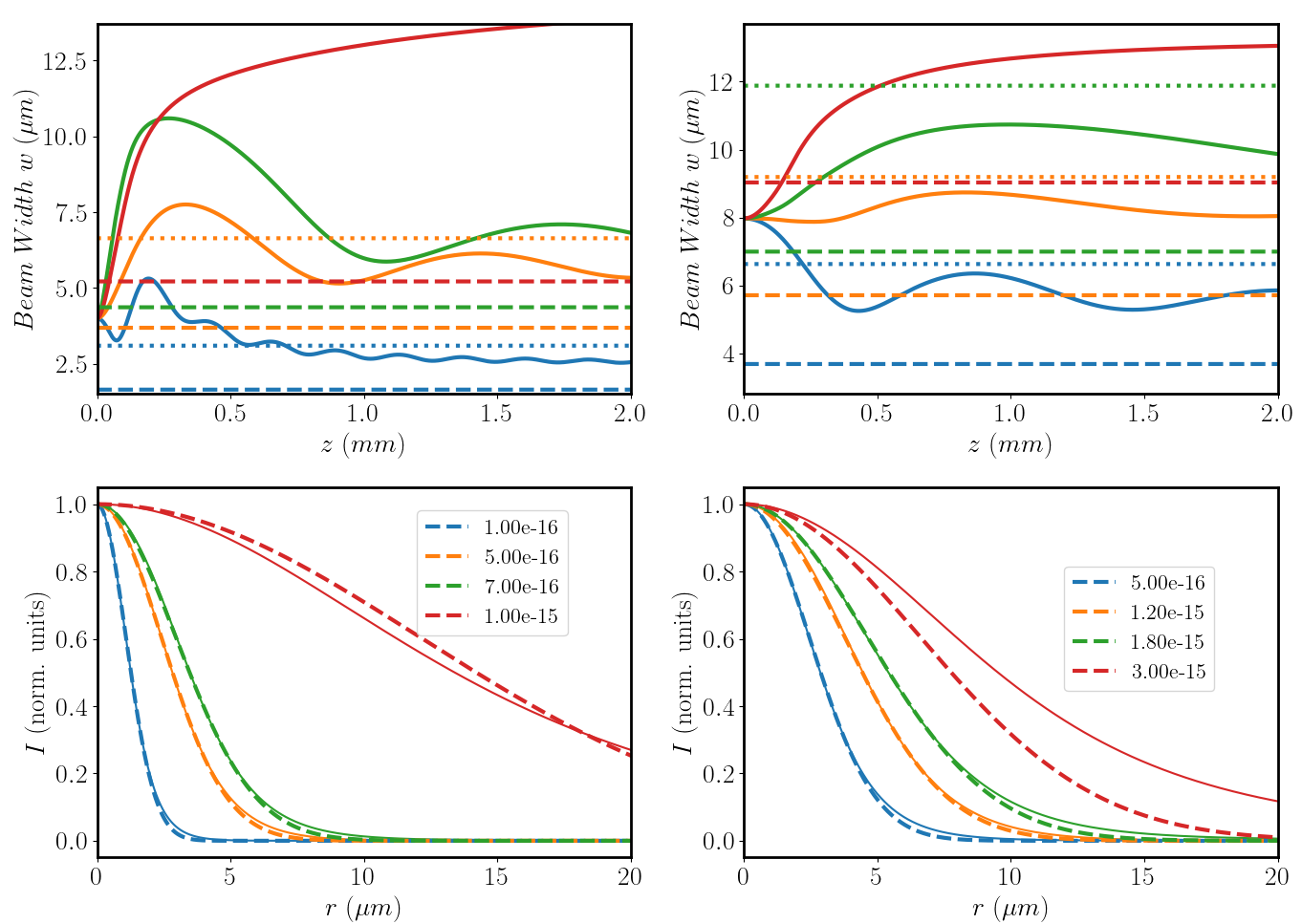}
    \caption{Beam width versus the propagation distance $z$ (top row) and output profiles versus the radial coordinate $r$ (bottom row) in cubic-quintic materials when the input width $w_0$ is $4~\mu$m (left column) and $8~\mu$m (right column). Each color corresponds to a different ratio $\eta^\prime$ between the two nonlinearities, see the legends in the bottom row. Dashed (dotted) straight lines in the top row show the minimum (average) soliton  width as defined by Eq.~\eqref{eq:min_wsol} (Eq.~\eqref{eq:wav_theory}). Dashed lines in the bottom row are Gaussian best-fits calculated by fixing the amplitude and varying only the width. In all the simulations shown in this figure the input power is fixed such that $P=2P_\mathrm{cr}$. Here $\lambda=1064~$nm, $N_0=2$, and $n_2=5\times 10^{-18}$m$^{2}$W$^{-1}$.} 
    \label{fig:simulations_vs_eta}
\end{figure}
Top row in Fig.~\ref{fig:simulations_vs_eta} shows the beam width evolution $w$ along $z$ as the amplitude of the quintic nonlinearity is varied. The power is fixed to $2P_\mathrm{cr}$, that is, when the Gaussian soliton ceases to exist according to the theory. The numerical results (solid lines) are compared with the average value $w_\mathrm{av}(P,w_0)$ provided by Eq.~\eqref{eq:wav_theory} and the minimum soliton width calculated from theory $w^\mathrm{min}_\mathrm{sol}$, see Eq.~\eqref{eq:min_wsol}. The agreement is quite good: once the stationary regime is achieved, the soliton width always remains above the theoretical limit set by Eq.~\eqref{eq:min_wsol}.
Breathing is minimized when the overlap between the input beam and the stationary solution is maximized. 
Regardless of the input width $w_0$, the average width tends to a value close to Eq.~\eqref{eq:wav_theory}, after a transition zone where the beam profile adapts to the stationary value \cite{Wright:1995}. Theoretical predictions for $w_\mathrm{av}$ are more accurate in the case of small breathing oscillations, although a slightly weaker mean confinement than numerical one is predicted even when breathing is minimal. Bottom row shows the corresponding intensity profiles sampled in $z=2~$mm. Whenever a condition close to equilibrium is achieved (that is, for not too large $|\eta|$ in the case shown ), a near-Gaussian shape is achieved, indeed confirming \textit{ex post} the applicability of our theoretical method. 

\subsection{Monodimensional Kerr media}

In a Kerr material propagating in one transverse direction and after defining the power per unit length $\mathcal{P}$, Eq.~\eqref{eq:ansatz_a} reads $a=\gamma_\mathrm{1D} \mathcal{P}/w^{2M}$; we can then set $N=1$. From Eq.~\eqref{eq:a_eff} we calculate $\gamma_\mathrm{1D} = \sqrt{2\pi}n_2$ and $M=1$. Therefore, the effective potential $U$ in (1+1)D Kerr case takes the same form of the HNL limit in the (2+1)D case  \cite{Alberucci:2016_1}. Furthermore, constant $B$ turns out to be independent of the initial beam width $w_\mathrm{in}$, %$B= 8/(k_0n_0 w_0)^2+ 4\gamma_\mathrm{1D} \mathcal{P}/(k_0 n_0 R_\mathrm{in} ) - 2\gamma_\mathrm{1D} \mathcal{P}/n_0 $. 
$B= 8/(k_0n_0 w_0)^2 - 2\chi  \gamma_\mathrm{1D} \mathcal{P}/n_0 $. The potential $U$ is then
\begin{equation}
    U(w^2) =  %-\frac{2\gamma_\mathrm{1D} P^N }{n_0} \frac{1}{w^{2(M-2)}} - Bw^2
     \frac{\sqrt{2\pi}n_2 \mathcal{P}}{8n_0} w^2 \left( \ln\frac{w^2}{w^2_\mathrm{av}} - C \right),
\end{equation}
where $C= 1- 8Bn_0/(\sqrt{2\pi}n_2 \mathcal{P}) $ and %$w_\mathrm{sol} = \left[4\pi^2/\left(k_0^2 n_0 n_2 \mathcal{P} \right) \right]^{1/2}$ 
$w_\mathrm{sol} = \left[4/\left( \sqrt{2\pi}k_0^2 n_0 n_2 \mathcal{P} \right) \right]^{1/2}$. Given that $\lim_{x\rightarrow \infty} x\left(\ln x - const \right) = + \infty $, the potential correctly predicts the existence of self-trapping, no matter what the input conditions are. Physically speaking, the HNL case in (2+1)D and the NLSE in (1+1)D share the same shape for the potential $U$ due to the existence of at least one soliton for any value of the input power \cite{Kivshar:2003}.
%\cite{Burak:1994}

\subsection{Nonlocal media}

\begin{figure}[h]
    \centering
  \includegraphics[width=0.99\linewidth]{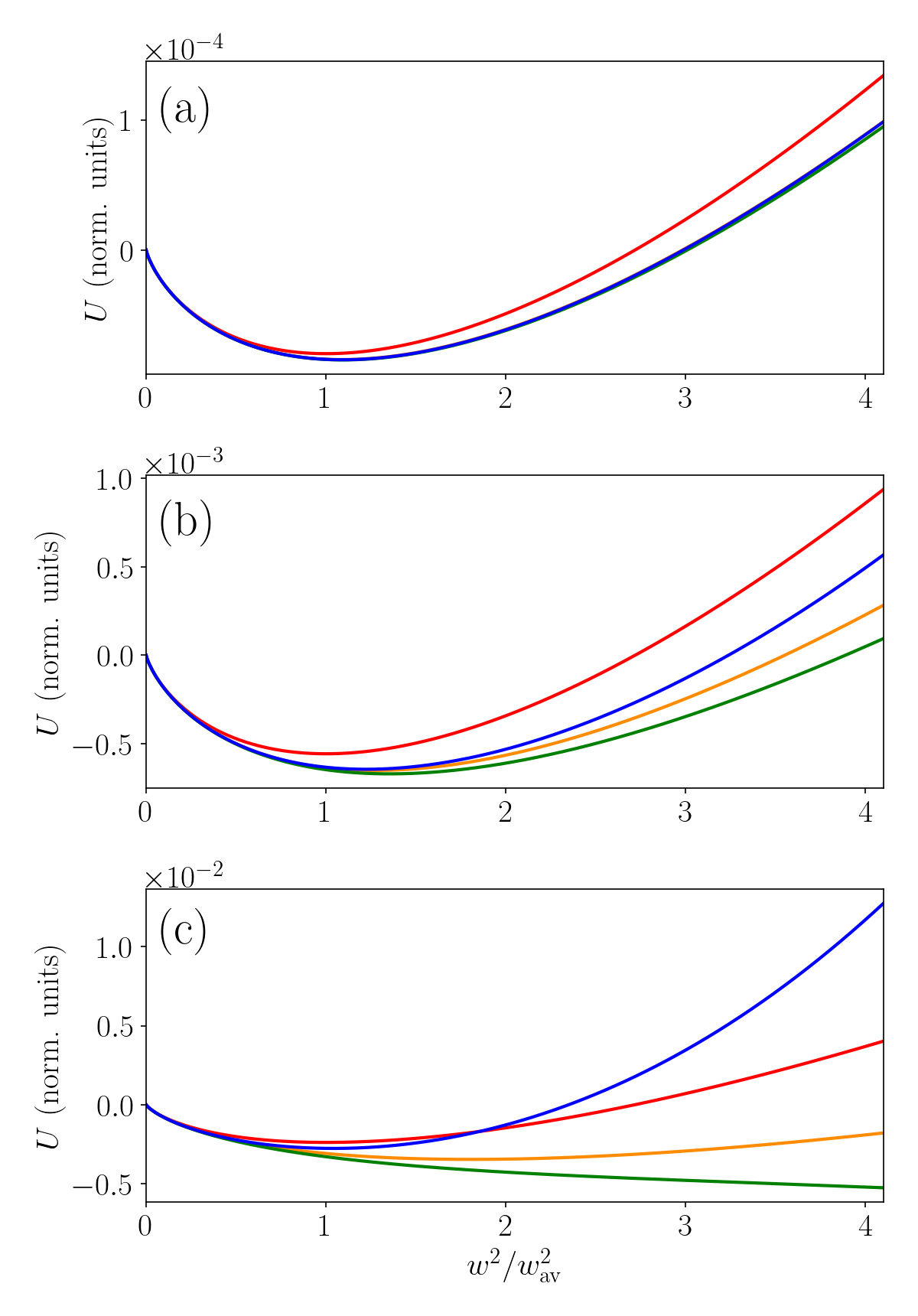}
    \caption{Profile of the potential $U$ expressed by Eq.~\eqref{eq:u_nonlocal} versus the squared beam width $w^2$ for $(w_\mathrm{av}/l)^2=1\times 10^{-3}$ (a), $7\times 10^{-3}$ (b) and $3\times 10^{-2}$ (c). Red curves are the potential in the HNL case, whereas orange, green and blue curves are computed including terms up to $m=-1,\ 0,\ 1$, respectively. }
    \label{fig:nonlocal_potential}
\end{figure}

Whereas Eq.~\eqref{eq:a_nonlocal_general} establishes the role played by nonlocality in self-trapping in its more general form, it is of difficult application in real cases due to the complexity associated with the series expansion of the Green function, plus the need to re-calculate $\sigma$ at each power. To simplify the terms of the problem, we assume a parabolic potential in a circle of radius $\sqrt{2}w$ around the origin. Such a choice ensures that the whole beam overlaps with the parabolic part of the nonlinear potential. This semi-empirical approach is used because an analog of Eq.~\eqref{eq:a_eff} in the nonlocal case would generate cumbersome and hard-to-interpret formulae. Outside this inner circle, we suppose the nonlinear potential to be $V_\mathrm{NL}=n_\mathrm{NL}PG(r,0)/k_0$. The QHO strength $a$ is then calculated by imposing continuity of the first derivative of the potential at the border between the two regions \cite{Conti:2003,Alberucci:2014}
\begin{equation} \label{eq:a_nonlocal}
    a= \frac{n_\mathrm{NL} P}{\sqrt{2} w} \left.\frac{d G(r,0)}{d r}\right|_{r=\sqrt{2} w}.
\end{equation}
 To fix the ideas, we consider a cylinder of radius much larger than $l$. In this limit we can assume an infinitely extended sample: the Green function then reads $G(r)=-\frac{1}{2\pi} K_0\left(\frac{\pi r}{l} \right)$, where $K_0$ is the modified Bessel function of second kind of order 0. The potential expressed as a power series of $\left(w/l \right)$ is
\begin{align} \label{eq:u_nonlocal}
    U &= \frac{n_\mathrm{NL} P}{ l^2}  \left( \sum_{m=-1}^{\infty} \frac{a_{m}}{\sqrt{2}n_0 \left( 1 + m/2 \right)} {\left(\frac{w}{l}\right)^{m+4}}\right. +  \nonumber \\
    & \left.  \frac{1}{4\pi}\left(\frac{w}{l} \right)^2 \left[\ln\left(\frac{w^2}{w^2_\mathrm{av}} \right) -1 \right] \right),
\end{align}
where the coefficients $a_m$ are calculated in Appendix~\ref{sec:V0_expansion}. With respect to the local case and multiple nonlinearities, all the terms in $U$ are directly proportional to the power $P$, as it should be due to the linear nature of Eq.~\eqref{eq:diffusion} with respect to the beam intensity. The power enters into play by determining the average width $w_\mathrm{av}$, whose dependency from $P$ is reported after Eq.~\eqref{eq:particle_potential_HNL}. The dynamics of the beam width then depends only on the ratio $w/l$: as the soliton broadens (that is, lower power $P$), more and more terms of the series needs to be accounted for. 
Furthermore, we have $U\propto l^{-2}$, showing how the amplitude of the nonlinear perturbation increases as the material response becomes more local. 
 Beyond the amplitude of the nonlinear perturbation, nonlocality strongly impacts both stability and breathing behavior of the self-localized beams.
 To analyze the role of nonlocality we plot the potential $U$ with terms up to $(w/l)^{5/2}$ (corresponding to $N=-1/2$) for different $w_\mathrm{av}/l$ in Fig.~\ref{fig:nonlocal_potential}. For $N=1$ the potential features a local minimum: the effective particle in the general case oscillates around the minimum, corresponding to a periodic breathing soliton. Neglecting the series in Eq.~\eqref{eq:u_nonlocal} works fine when $\left(w_\mathrm{av}/l\right)^2<1\times 10^{-5}$, that is, $w_\mathrm{av}/l<3.2\times 10^{-3}$. The potential $U$ computed by halting the series at a finite number of terms (up to three terms, corresponding to $m=1$ or equivalently $N=1/2$) is plotted in Fig.~\ref{fig:nonlocal_potential} for three different $w_\mathrm{av}/l$. First thing to notice, the accuracy of the approximation decreases as $w/l$ grows, in agreement with the form of the series expansion. When $\left(w_\mathrm{av}/l\right)^2=1\times 10^{-3}$ [Fig.~\ref{fig:nonlocal_potential}(a)], the potential $U$ shifts towards the right side and slightly downwards, with the component $N=1/2$ ($m=-1$ in Eq.~\eqref{eq:u_nonlocal}) providing the dominant contribution. When $\left(w_\mathrm{av}/l\right)^2=7\times 10^{-3}$ [Fig.~\ref{fig:nonlocal_potential}(b)], higher order contributions becomes relevant. Despite that, the overall potential conserves its shape, that is, the presence of a local minimum corresponding to a stable spatial soliton, whose width becomes wider as $w_\mathrm{av}/l$ gets larger.  When $\left(w_\mathrm{av}/l\right)^2=3\times 10^{-2}$ [Fig.~\ref{fig:nonlocal_potential}(c)], the large differences between the curves show how many additional terms are required for achieving a correct approximation of the full potential. Summarizing, at very high power $w_\mathrm{av}/l \ll 1$, the self-confinement follows the HNL case established by Eq.~\eqref{eq:particle_potential_HNL}. At lower powers, the soliton width gets wider than in the HNL case. When $w_\mathrm{av}/l$ approaches unity, the soliton width rapidly increases. From a physical point of view, the system is approaching the local limit where soliton width is independent from the power $P$.  The associated critical power can be found substituting into Eq.~\eqref{eq:Pcr} the effective Kerr coefficient extracted from Eq.~\eqref{eq:VNL_local_limit}, in turn providing in the nonlocal case $P_\mathrm{cr}= \pi\lambda^2 /\left( 2\sigma_1 n_0 n_\mathrm{NL} l^2 \right)$. Stable solitons cannot carry a power larger than the critical value: therefore, a vertical asymptote is expected on the left side (i.e., at lower powers self-confinement is inhibited because the soliton is approaching the local limit) of  the soliton existence curve in the plane power-width. In fact, $P_\mathrm{cr}\rightarrow \infty$ in the local limit $l\rightarrow 0$. In terms of effective potential, wherever $P>P_\mathrm{cr}$ a local minimum in $U$ is present, whereas for powers below this threshold $U$ is monotonically decreasing, when plotted versus the normalized width $w/l$. The described dynamics is in agreement with the theoretical and numerical results reported in Figs.~4,5 in Ref.~\cite{Alberucci:2014}.

%\section{Weakly nonlocal media}
%\subsection{Saturable media}

%\cite{Lam:1975}
%\cite{Marburger:1975}

\section{Conclusions}
\label{sec:conclusions}
%Variational for collapse \cite{Desaix:1991}.   

We discussed a new theoretical approach to the modeling of self-focusing and self-trapping based upon the Ehrenfest theorem. We showed how the dynamics of the beam width can be modeled with good accuracy as a quantum harmonic oscillator whose strength depends both on the input power and beam width. %We showcased the versatility of our approach, capable of describing the wave propagation in the presence of Kerr-like nonlinearities, regardless of their nonlocality degree. 
As a direct consequence of using the Ehrenfest's theorem as a starting point, in our approach the fundamental quantity to model the wave evolution is the second-moment $w^2$. The dynamics is then determined by the convexity of $w^2$, whose local changes are determined by the nonlinear effects at work.
Therefore, analogously to the variational and moments method, our approach in the end provides a 1D effective potential with power-dependent features, thus reducing the nonlinear wave model to a standard mechanical problem. 
Our method explicitly and clearly shows that the input conditions do not solely affect the boundary conditions, but instead they alter the effective mechanical potential itself. \\
By comparison with the available literature and numerical simulations, in a second step we demonstrated the versatility of our approach, capable of providing at the same time an accurate description and intuitive picture of nonlinear propagation in several different types of nonlinear media, including pure Kerr, cubic-quintic, and nonlocal nonlinearities. We showed how each nonlinear mechanism yields an effective potential consisting in a sum of terms dependent on the powers of the second-moment $w^2$; importantly, the potentials can be superposed when multiple nonlinearities are simultaneously acting on the wave. In the Kerr case, our approach permits to achieve closed-form solutions for the features of the nonlinear Gaussian beams, both before and after the critical power, and independently from the beam size. As a potential application, our results pave the way to new experimental techniques to measure the Kerr coefficient using the beam dynamics before the catastrophic collapse. Additionally, a straightforward generalization of our approach can model the interplay between nonlinearities featuring different degrees of nonlocality. Another interesting generalization is the case when the wave does not fulfill cylindrical symmetry on the transverse plane. 

In the current work we limited our attention to the case of lossless CW focusing. A first step forward would be to generalize our model to the case of nonlinear losses in the monochromatic regime \cite{Couairon:2003,Falcao:2013}. More interesting is the application to the case of ultrashort pulses \cite{Couairon:2007,Gattass:2008,Lim:2014} -of great relevance in real experiments and applications- along two different directions, both of them related to a proper accounting for the role played by the temporal profile of the pulse. On one side, the effect of temporal dispersion can be accounted by adding a third transverse coordinate \cite{Rothenberg:1992,Skanka:1997,Mlejnek:1998}. On the other side, plasma nonlinearity plays a fundamental role in halting the catastrophic collapse of ultrashort pulses \cite{Polynkin:2013,Alberucci:2026}, but it introduces temporal nonlocal effects due to the generation of excited electrons with a lifetime longer or comparable with the pulse duration itself.

\section*{Acknowledgments}

This work is supported by the Free State of Thuringia and the European Social Fund Plus (2022FGR0002). We acknowledge the financial support of Deutsche Forschungsgemeinschaft (DFG) through the Collaborative Research Center CRC 1375-NOA (Nonlinear Optics down to Atomic scale).
 
\appendix

\section{Ruling equation in the presence of non-adiabatic losses}
\label{sec:der_power}
When losses are not adiabatic, the derivatives of the power versus $z$ need to be accounted for. Equation~\eqref{eq:equation_fourth_order} becomes
\begin{align} 
    &\frac{d^2}{dz^2} \left[ \frac{d^2 w^2}{dz^2} + \frac{2\gamma P^N (N-2)}{n_0 (N-1)} \frac{1}{w^{2(N-1)}} \right] = \nonumber \\ &-\frac{\gamma N P^N}{w^{2(N-1)}}\left[\frac{3}{w^2}\frac{dw^2}{dz}\frac{d\ln P}{dz} + \frac{d^2\ln P}{dz^2} + N \left(\frac{d\ln P}{dz} \right)^2 \right].
\end{align}
In the case of linear losses $P\propto e^{-\alpha_\mathrm{loss} z}$, $\frac{d\ln P}{dz} = \alpha_\mathrm{loss}$. Thus, dissipation affects can strongly modify the dynamics of self-trapping, in agreement with what has been demonstrated in nearly integrable systems \cite{Kivshar:1989}.

\section{Tight and deep focusing in the Kerr case}
\label{sec:tight_focusing}

From Eq.~\eqref{eq:zc} and $w^2_\mathrm{0NL}\approx(1-P/P_\mathrm{cr})w^2_0$, the nonlinear shift $z_c-z_\mathrm{0NL}$ for $P>P_\mathrm{cr}$ and tight focusing can be recast as
\begin{equation}
    z_c-z_\mathrm{0NL} \approx - \frac{n_0 k_0 w^2_0}{2} \sqrt{\frac{P}{P_\mathrm{cr}}-1}.
\end{equation}
Accordingly, the dashed lines in Fig.~\ref{fig:NLGB_parameters}(a) follow a square root trend with respect to power. Furthermore, they are independent of $z_0$ (compare black and red lines), whereas their amplitude depends quadratically on the linear waist $w_0$ (compare the blue and black lines). \\
Substituting Eq.~\eqref{eq:alpha} into Eq.~\eqref{eq:NLGB} evaluated at $z=0$ and remembering again that $w^2_\mathrm{0NL}=(1-P/P_\mathrm{cr})w^2_0$, the nonlinear focal position reads
\begin{equation}
    z_\mathrm{0NL} = \frac{n_0 k_0 w_0 w_\mathrm{in}}{2} \sqrt{1- \left(1 -\frac{P}{P_\mathrm{cr}} \right)\frac{w^2_0}{w^2_\mathrm{in}}}.
\end{equation}
Rewriting $w_\mathrm{in}$ in terms of the linear focal position $z_0$
\begin{equation}  \label{eq:z0NL_appendix}
    z_\mathrm{0NL} = z_0 \sqrt{1+ \frac{n_0^2 k_0^2 w^4_0}{4z_0^2}\frac{P}{P_\mathrm{cr}}}.
\end{equation}
Let us now compare the predictions of Eq.~\eqref{eq:z0NL_appendix} with Fig.~\ref{fig:NLGB_parameters}(a). Linearization of Eq.~\eqref{eq:z0NL_appendix} is more accurate for smaller waists $w_0$ (compare the blue line with the red and black ones, the latter being linear in the range shown).
For the same $w_0$, the slope versus the power is steeper for shorter focal positions $z_0$ (compare the black line with the red one).

\section{Effective energy in the Kerr case}
\label{sec:energy_kerr}

The energy $E$ for the equivalent mechanical system reads $E=\frac{1}{2}\left(\frac{dw^2 }{dz} \right)^2- \frac{8 }{n_0^2 k_0^2 w_{0}^2}\left(1-\frac{P_\mathrm{cr}}{P^*_\mathrm{cr}} \right)w^2$. After accounting for the initial condition for the beam width $w(z=0)=w_\mathrm{in}$ and the pseudo-velocity $\frac{dw^2}{dz}$ [see Eq.~\eqref{eq:initial_curvature_radius}] we find
\begin{align} \label{eq:energy_kerr_case}
    %E = \frac{2w^4_\mathrm{in}}{R^2_\mathrm{in}} - \left( 1- \frac{P}{P_\mathrm{cr}} \frac{w_0^2}{w^2_\mathrm{in}} \right) w^2_\mathrm{in}.
E = \frac{\beta^2 }{2}\left(\frac{P}{P^*_\mathrm{cr}} \right)^2 + w^2_\mathrm{in}\left(\frac{8}{k_0^2 n_0^2 w_0^2}-\frac{2\beta}{R_\mathrm{0}} \right)\frac{P}{P^*_\mathrm{cr}} \nonumber \\
+ w^2_\mathrm{in} \left(\frac{2w^2_\mathrm{in}}{R^2_\mathrm{0}}-\frac{8}{k_0^2 n_0^2 w_0^2} \right),
\end{align}
where we defined $\beta = \frac{2w^2_0}{k_0 L^2}= 8/(k_0^3 n_0^2 w_0^2)$. After straightforward computation we find the more compact expression
%\begin{equation}
%    E = \frac{8}{k_0^2 n_0^2} \left[  \frac{4}{\chi^2 n_0^2 k_0^4 w_\mathrm{in}^4} \left(\frac{P}{P_\mathrm{cr}} \right)^2+ \frac{P}{P_\mathrm{cr}} - 1 \right]
%\end{equation}
\begin{equation}
    E = \frac{8}{k_0^2 n_0^2} \left[  \Pi \frac{w_0^4}{w_\mathrm{in}^4} \left(\frac{P}{P_\mathrm{cr}} \right)^2 + \Sigma\left(\chi\right) \frac{P}{P_\mathrm{cr}} + \Omega\left(\chi\right) - 1 \right], \label{eq:energy_suppl}
\end{equation}
where $\Pi = \frac{4}{n_0^4 k_0^4 w^4_0}$, $\Sigma = \left( 2\chi^2 + \chi -1 \right)/(2\chi)$ and $\Omega=\left(k_0^4 w_\mathrm{in}^4 n_0^2 \right)\left( \chi -1\right)^2/\left(16 \chi^2\right)$. \\
Let us start to discuss the simplest case when the impinging beam is focused on the entrance facet, therefore $z_0=0$. For flat phase fronts at the input, it is $\chi=1$ and $w_\mathrm{in}=w_0$. Equation~\eqref{eq:energy_suppl} then yields
$E= \frac{8}{k_0^2 n_0^2}\left[ \Pi \left(\frac{P}{P_\mathrm{cr}} \right)^2+ \frac{P}{P_\mathrm{cr}} - 1   \right]$. 
For $P=P_\mathrm{cr}$ we can define $E=\Pi/(k_0 n_0)^2=E_0$, where $E_0$ corresponding to the kinetic energy associated with the self-focusing at the entrance interface.  Accordingly, $E_0$ decreases for broader inputs and for shorter wavelengths. From its definition, $E_0$ is negligible small in the scalar regime of the Maxwell's equation, and the quadratic term in Eq.~\eqref{eq:energy_suppl} with respect to $P$ can be safely neglected. Therefore, whenever $P<P_\mathrm{cr}$ ($P>P_\mathrm{cr}$), the effective energy $E$ is negative (positive). As a matter of fact, catastrophic collapse occurs only for excitations exceeding the critical power. %We also have  $(E-E_0)>0$ [$(E-E_0)<0$] when $P>P_\mathrm{cr}$ ( $P<P_\mathrm{cr}$).

%For flat phase fronts at the input $E= \frac{8}{k_0^2 n_0^2}\left(\frac{P}{P_\mathrm{cr}} -1 \right) + \frac{\beta^2 }{2}\left(\frac{P}{P_\mathrm{cr}} \right)^2$.
In the presence of an initial phase front, we showed in the main text that $ \chi \approx 1$. We then evince $\Sigma \approx 1$ and $\Omega \approx 0$. The sign of $E$ with respect to the input power $P$ is identical to the case of planar wavefronts discussed above. Thus, when $P_\mathrm{cr}<P<P^*_\mathrm{cr}$, energy is positive: the effective particle moves inside the region of negative widths, even if the potential energy $U$ is monotonically descending. 
%At the critical power $P=P_\mathrm{cr}$ it is $E=E_0  \frac{w_0^4}{w_\mathrm{in}^4}$: the initial kinetic energy 
 \\
%To achieve a more intuitive physical picture, we can recast Eq.~\eqref{eq:energy_kerr_case} in terms of the parameters $w_0$ and $z_0$ of the incident Gaussian beam
%\begin{equation}
 %   E = \frac{\beta^2 }{2}\left(\frac{P}{P_\mathrm{cr}} \right)^2 + w^2_\mathrm{in}\left(1-\frac{2\beta}{R_\mathrm{in}} \right)\frac{P}{P_\mathrm{cr}} + w^2_\mathrm{in} \left(\frac{2w^2_\mathrm{in}}{R_\mathrm{in}}-1 \right). %\left[\frac{P}{P_\mathrm{cr}} + \left( \frac{32z_0^2}{k_0^4 n_0^4 w_0^6} - \frac{4z_0^2}{k_0^2 n_0^2 w_0^4} -1 \right) \right]  w_0^2.
%\end{equation}

%\section{Non-planar input wavefront in cubic-quintic media}
%\label{sec:cubic_quintic}

%When $w_0\neq w_\mathrm{in}$, we can recast 
%Eq.~\eqref{eq:B_interplay} by rewriting $P/P_\mathrm{cr}$ as a Taylor series for small $z_0/L$
%\begin{equation}
%    B\approx B_0 + \frac{32 z_0^2}{n_0^4 k_0^4 w_0^6} \left(1+ \eta I_0^{N_0-1}\right) \frac{P}{P_\mathrm{cr}},
%\end{equation}
%where $B_0$ is the value of $B$ when $w_\mathrm{in}=w_0$, that is, planar wavefront at the input. Thus, small shifts of the focal point $z_0$ does not alter the overall shape of the potential~\eqref{eq:potential_focus_defocus}: the main effect is favoring self-confinement at large width $w$. 

\section{Numerical computation of the soliton shapes in cubic-quintic materials}
\label{sec:NL_eigenvalue}
\begin{figure}
    \centering
  \includegraphics[width=0.99\linewidth]{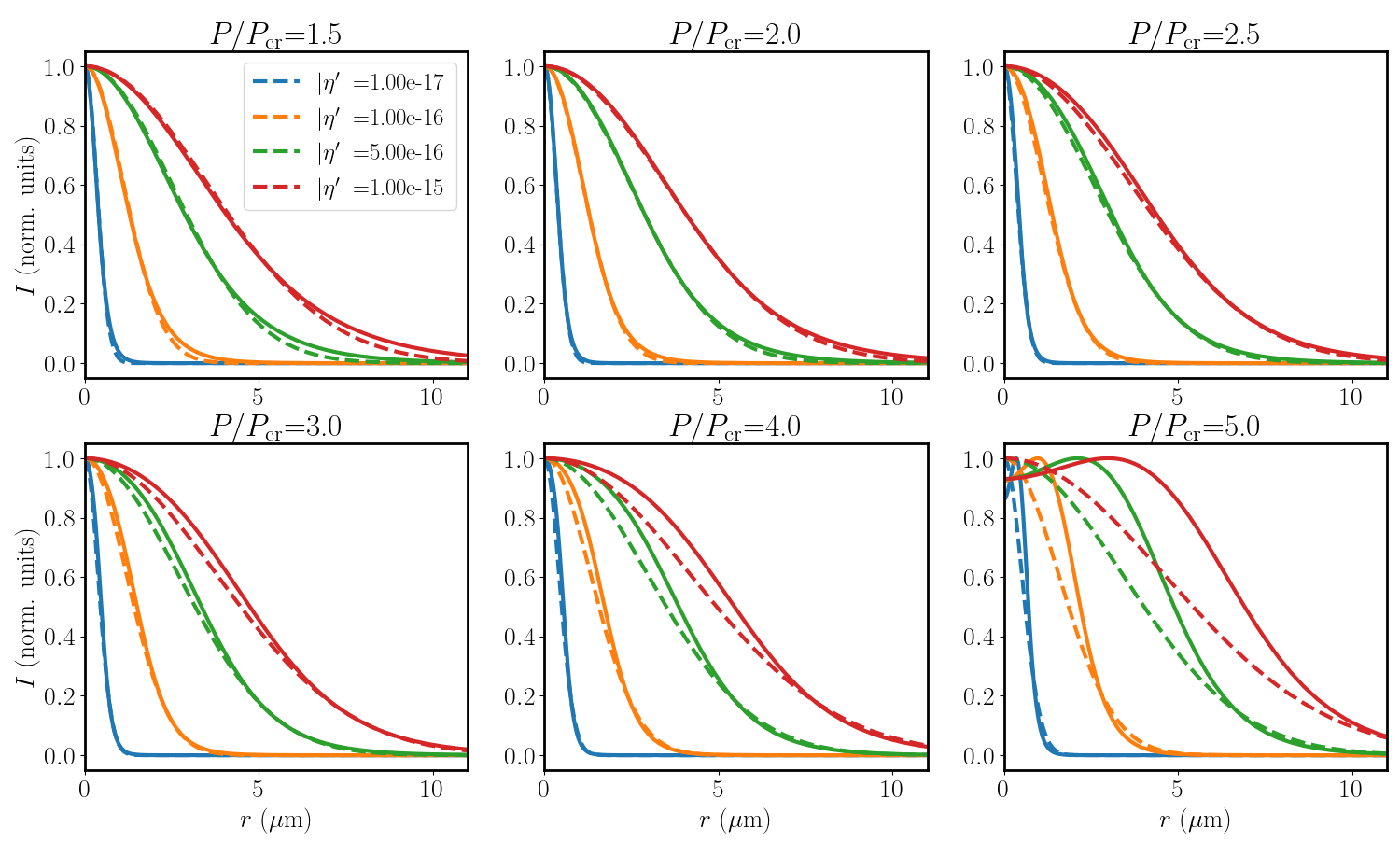}
    \caption{ Soliton profile versus the radial coordinate $r$. Solid and dashed lines are the numerical solution and the corresponding Gaussian best-fit, respectively; each color corresponds to a different ratio $\eta^\prime$ (see the legend in the first panel for the numeric values). Each panel shows a different power, see the title for the specific values. Here $\lambda=1064~$nm, $N_0=2$, and $n_2=5\times 10^{-18}$m$^{2}$W$^{-1}$. }
    \label{fig:soliton_numerics}
\end{figure}

The profile of the solitons for a given nonlinearity can be found by writing the NLSE in the form $\partial_z \psi = \hat{H}\psi$, discretizing the operator $\hat{H}$ using finite-differences, and finally solving the nonlinear eigenvalue problem. The dependence of $\hat{H}$ itself on the solution $\psi$ is managed by using an iterative algorithm. The lowest-order self-trapped wave is selected by looking for the eigenvalue closer from below to the maximum of the index well. Numerical results are shown in Fig.~\ref{fig:soliton_numerics}.  Regardless of the ratio $\eta$ between the two nonlinearities, a similar dynamics versus the normalized carried power $P/P_\mathrm{cr}$ is observed. Up to $P/P_\mathrm{cr}=2$, the Gaussian approximation is very accurate. For further increases in power, the concavity of the numerical profile near the origin becomes larger than the Gaussian best-fit. Physically, the defocusing effect starts to be relevant, effectively saturating the self-focusing in correspondence of the largest intensities. For $P/P_\mathrm{cr}\approx 4.6$, the defocusing becomes strong enough to flatten the soliton near the origin, eventually leading to the emergence of a local minimum in the beam center, the intensity peak lying now in a ring \cite{Wright:1995}.

\section{Dynamical simulations of the NLSE}
\label{sec:numerics}
\begin{figure}
    \centering
  \includegraphics[width=0.99\linewidth]{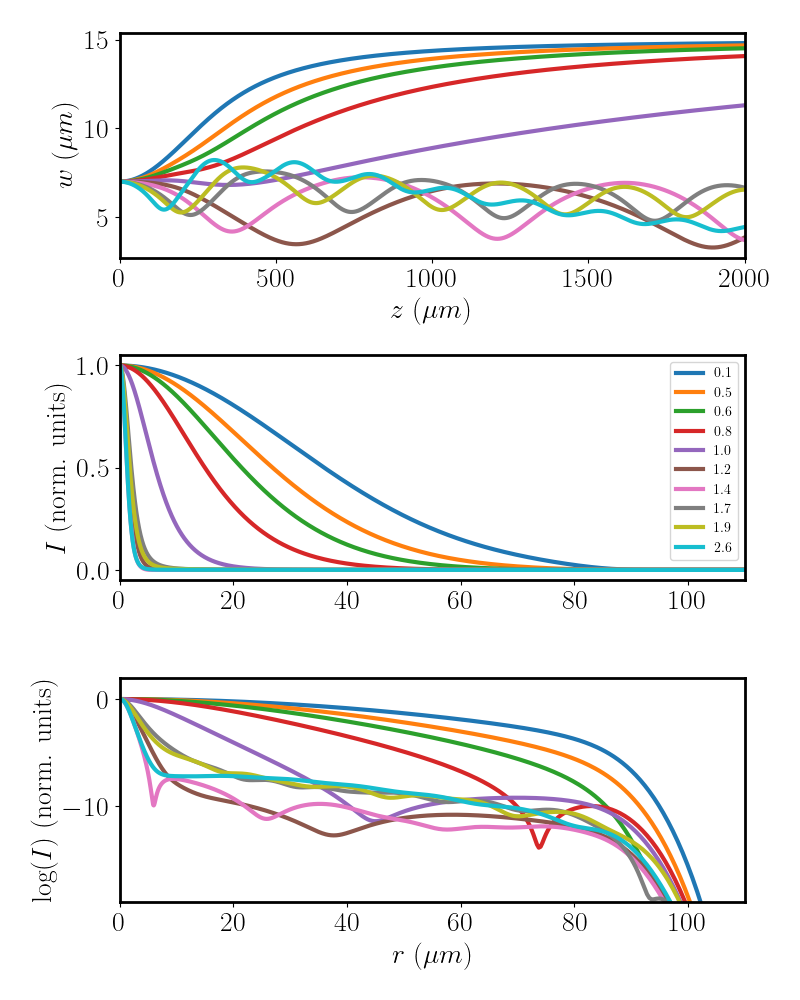}
    \caption{BPM simulations in a cubic-quintic medium when the input waist is $7~\mu$m and the impinging phase front is planar. In the top panel each line corresponds to the evolution versus $z$ of the beam width for a given input power. Corresponding intensity profile at $z=2~$mm in linear (middle panel) and log (bottom panel) scale. Correspondence between the normalized input power $P/P_\mathrm{cr}$ and colors is provided in the mid panel. Missing simulation parameters are provided in the text, see Appendix~\ref{sec:numerics}. }
    \label{fig:BPM_focus_defocus}
\end{figure}

The NLSE is simulated in radial coordinates assuming a cylindrically-symmetric propagation. A logpolar coordinate system is employed to improve the numerical accuracy. The optical propagation is simulated using an operator splitting, with the diffraction operator being simulated using an implicit Crank-Nicolson scheme \cite{Press:1992}. As boundary conditions at large radii $r$, we impose a vanishing field. Numerical back-reflection from the edges of the grid are damped out using a super-Gaussian attenuator. \\
Figure~\ref{fig:BPM_focus_defocus} shows -as an example- the beam width versus the propagation distance $z$ computed for different powers, represented as the ratio $P/P_\mathrm{cr}$ in the associated legend.  The material nonlinear response is cubic-quintic, that is, $N=1$ for the focusing Kerr response and $N=2$ for the higher order defocusing nonlinearity. The material parameters are $n_2= 5\times 10^{-18}$m$^2$W$^{-1}$ and $n_4= - n_2\times 10^{-17}$m$^2$W$^{-1}$, whereas $n_0=1.5$ and $\lambda = 1064~$nm. The impinging beam is assumed having a flat wavefront. %Different colors correspond to a different input waist. 
Due to the presence of radiation at large transverse positions coming from the first stage of propagation \cite{Malkin:1993,Wright:1995}, the beam width is computed over a finite interval; in radial coordinates and in the cylindrical symmetric case we find $\langle x^2 \rangle = \sqrt{2\int_0^{r_0} I r^3 dr/\sqrt{\int_0^{r_0}Irdr}} $. In the results shown we have set $r_0 = 5~\mu$m. With respect to the beam width, a clear light localization in the form of a breather is observed for $P \geq 1.2 P_\mathrm{cr}$, whereas self-focusing takes place for lower powers. From the output profile it is evident that a broad solitary wave is also excited at the critical power, but it is not appreciated from the width evolution due to our windowing. For $P>2.6P_\mathrm{cr}$ our code for the chosen numerical grid starts to encompass numerical instability (i.e., the intensity diverging in the origin) at long enough propagation distances. 

\section{Power series expansion for the QHO strength in the nonlocal case}

\label{sec:V0_expansion}

 For shape-preserving solitons, the general solution of Eq.~\eqref{eq:diffusion} can be written in terms of the 2D Green function $G(\bm r_T,\bm r_T^\prime)$, $V_\mathrm{NL}(x,y) =  n_\mathrm{NL}\int{I(\bm r_T^\prime)G(\bm r_T,\bm r_T^\prime)dx^\prime dy^\prime} /k_0$ %=\frac{2  n_\mathrm{NL} P}{\pi w^2}\int{e^{-\frac{2r^2}{w^2}}G(\bm r_T,\bm r_T^\prime)dx^\prime dy^\prime}$. 
 The Green function of a 2D Poisson equation is singular for $\bm r_T=\bm r_T^\prime$, thus we need to perform the integral before computing the value or any derivative of the nonlinear potential $V_\mathrm{NL}$ in the origin $(x=0,y=0)$. This problem can be circumnavigated defining a nonlinear potential parabolic around the origin, whereas sharing the shape of the Green function computed at the origin (i.e., $\bm r_T^\prime=0$) on the tails; for this model, the transition between the different shapes occurs in $r^*=\sigma w$, $\sigma$ being dependent on the soliton power. Within this approximation, the value in the origin $r=0$ of the nonlinear perturbation $V_0$ is $V_0 \approx  n_\mathrm{NL} P \left( G^* + \sigma^2/(2\pi) \right)\left/\left( 1 + \frac{1}{4}\left(\frac{\pi \sigma w}{l}\right)^2\right)\right.$, where we set $G^* = G(r^*,0)$. \cite{Alberucci:2014} Expanding the Green function as $G=\left.\left(\sum_j G_j r^j\right)\right/r$, we finally find the series expansion for $V_0$ versus the beam width $w$
 \begin{align} \label{eq:V0_NL_power_expansion}
     V_0  = & n_\mathrm{NL} P \left( \frac{G^*_0}{\sigma w} + G^*_1 + \frac{\sigma^2}{2\pi}   + \sigma G^*_2 w + \sigma^2 G^*_3 w^2 + o(w^2) \right) \nonumber \\ & \times \left( 1- \frac{1}{4}\left(\frac{\pi \sigma w}{l}\right)^2 + o(w^2) \right).
 \end{align}

\section{Series expansion of the nonlocal Green function in the case of an infinite cylinder}

Given that $K^\prime_0(\frac{\pi r}{l})= -\frac{\pi}{l} K_1(\frac{\pi r}{l})$, we need to expand $K_1$ in its power series. In Ref.~\cite{Molu:2017} it has been shown that $K_1$ can be accurately approximated by the following finite sum  
\begin{equation}
    K_1\left(\frac{\pi r}{l}\right) \approx \frac{e^{-\frac{\pi r}{l}}}{\frac{\pi r}{l}} \sum_{q=0}^k{p_{1,k,q} \left(\frac{\pi r}{l}\right)^q}.
\end{equation}
In the previous formula we introduced the quantities  $p_{1,k,q} = \sum_{m=q}^k \Lambda(1,m,q)$ where 
\begin{equation}
     \Lambda(1,m,q) = \frac{(-1)^q\sqrt{\pi} \Gamma\left( m-0.5\right) L(m,q)} {2^{-q } \Gamma(-0.5)\Gamma\left( m+1.5\right) m! }.
\end{equation}
$\Gamma$ is the Gamma function and $L$ are the Lah numbers.
Making use of Table 1 in Ref.~\cite{Molu:2017}, we can write different expressions for $K_1$ according to the maximum degree of the polynomial used for the approximation
\begin{align}
     & K_1 \approx \frac{e^{-x}}{x} \left( 
1 + 0.8 x -0.1333 x^2 \right), \nonumber \\
 & K_1 \approx \frac{e^{-x}}{x} \left( 
1 + 0.8571 x -0.2476 x^2 + 0.0381 x^3 \right), \nonumber \\
 & K_1 \approx \frac{e^{-x}}{x} \left( 
1 + 0.8889 x -0.3429 x^2 + 0.1016 x^3 -0.0106x^4 \right), \nonumber
\end{align}
where we have set $x=\frac{\pi r}{l}$ for sake of compactness. 
For $a$ we finally find the following expression up to the linear term in $w/l$
\begin{equation}
    a = \frac{n_\mathrm{NL}P}{2\sqrt{2} l^2} \left[ a_{-2}\left(\frac{l}{w}\right)^2  + a_{-1}\frac{l}{w} + a_0 + a_1 \frac{w}{l}  \right],
\end{equation}
where $a_{-2}=\left(\pi \sqrt{2}\right)^{-1}$, $a_{-1} = -0.1429$, $a_0= -0.4865 $, $a_1=10.8$.  The effective particle potential $U$ at this level of approximation will then comprise four distinct terms, corresponding to $N=1,1/2,0,-1/2$. As the material goes more and more local (growing $w/l$), more terms will become relevant in determining $a$. Such an expression is in agreement with the general form provided by Eq.~\eqref{eq:V0_NL_power_expansion}.

%\section{Details about the FDTD simulations}

\bibliography{apssamp}% Produces the bibliography via BibTeX.

\end{document}